\newif\ifAPPENDIX \APPENDIXtrue

\newif\ifHYPERBORDER \HYPERBORDERfalse

\documentclass[runningheads]{llncs}

\usepackage{cmap}

\usepackage{amsmath}
\usepackage{amssymb}
\usepackage[only,llbracket,rrbracket,rightarrowtriangle]{stmaryrd} 
\usepackage{mathtools}
\usepackage{scalerel}    
\usepackage{stackengine} 

\usepackage{tikz-cd}

\usepackage[inline,shortlabels]{enumitem}

\usepackage[T1]{fontenc}
\usepackage{graphicx}
\usepackage{xcolor}

\usepackage{hyperref}

\ifHYPERBORDER
  \hypersetup{
    linkbordercolor=orange!70!black,
    citebordercolor=purple!80!blue,
    pdfborderstyle={/S/D/D[1.5 1]/W 0.333} 
  }
\else
  \hypersetup{
    colorlinks = true,
    linkcolor = gray,
    citecolor = purple,
    urlcolor=blue,
  }
\fi

\usepackage[style=lncs]{biblatex}
\makeatletter
\newcommand\xlabel[2][]{{\phantomsection\def\@currentlabel{#1}\label{#2}}}
\makeatother

\usepackage{mnsymbol-font}

\usepackage[scr]{rsfso}

\newcommand\In {\mathpunct{:}} 
\newcommand\Fun {\mathbin{\rightarrow}} 
\newcommand\Iff {\mathrel{\Longleftrightarrow}} 
\newcommand\Imp {\ensuremath\Rightarrow}

\newcommand\IMP {\ensuremath{\mathrel{\Longrightarrow}}}

\newcommand\ProgFromTo {\mathbin{\rightarrowtriangle}}

\DeclarePairedDelimiterX\setcmp[2]\{\}%
    {#1 \mathclose{}\nonscript\;\delimsize|\nonscript\;\mathopen{} #2}

\newcommand\Powerset {\ensuremath{\mathbb{P}}}
\newcommand\PowersetNonempty {\ensuremath{\mathord{\mathbb{P}\!_{%
    \ThisStyle{\raisebox{2\LMpt}{$\SavedStyle\scaleobj{0.7}{+ \mkern -1mu}$}}%
}}}%
}
\newcommand\Bool {\mathsf{Bool}}

\newcommand\LeastFixedPoint {\mathop\mathsf{LFP}}
\newcommand\Endo {\mathop\mathrm{Endo}}

\newcommand{\Kl} {\ensuremath{\mathrm{Kl}}}

\newcommand\False {\mathtt{false}}
\newcommand\True {\mathtt{true}}

\newcommand\Nat {\mathbb{N}}
\newcommand\Ints {\mathbb{Z}}
\def\RExtd{\ensuremath{\mathord{\bar{\mathbb{R}}_+}}}

\newcommand\Terminates {\ensuremath{\top}}

\makeatletter
\newdimen\@supscriptdim 
\newcommand\dirjoin[1][] {\ensuremath{
  \settowidth{\@supscriptdim}{$\m@th ^{\smash\uparrow}$}
  \mathop{\bigvee\nolimits^{\ifblank{#1}{}{\mathrlap}{\smash\uparrow}}}%
  \ifblank{#1}{}{_{#1}\hspace{\@supscriptdim}}
  }
}

\makeatother

\DeclareMathSymbol{\tbigsqcap}{\mathop}{mnsymbols}{"28}
\DeclareMathSymbol{\dbigsqcap}{\mathop}{mnsymbols}{"29}

\def\AA{\ensuremath{\mathcal{A}}}
\def\CC{\ensuremath{\mathcal{C}}}

\def\HH{\ensuremath{\mathcal{H}}}
\def\SS{\ensuremath{\mathcal{S}}}

\def\OO{\ensuremath{\mathcal{O}}}
\def\VV{\ensuremath{\mathcal{V}}}
\def\WW{\ensuremath{\mathcal{W}}}
\def\XX{\ensuremath{\mathcal{X}}}
\def\YY{\ensuremath{\mathcal{Y}}}
\def\ZZ{\ensuremath{\mathcal{Z}}}

\def\Obsvs{\mathrm{\Omega}}

\def\shorttimes{%
  \edef\theSpace{\mspace{0mu plus 1.5mu minus 0.75mu}}%
  \ensuremath{\theSpace\mathord\times\theSpace}%
}

\def\BB{\ensuremath{\mathcal{B}}}

\def\CtxP{\ensuremath{\mathord{\mathscr{P}\mkern-2mu}}}
\def\CtxQ{\ensuremath{\mathord{\mathcal{Q}}}}

\newcommand\Pred{\ensuremath{\mathbb{L}}}
\newcommand\Loss{\ensuremath{\mathbb{S} \Pred}}

\newcommand\Dist{\ensuremath{\mathbb D}}
\newcommand\DistPartial{\ensuremath{\mathord{{\mathbb D}_{\scaleobj{0.7}{\!\bot}}}}}

\newcommand\supp {\ensuremath{\mathop{\mathrm{supp}}}}

\newcommand\PC[1] {\ensuremath{\mathbin{\!_{\smash{#1}\!}\oplus}}}

\def\IB[1]{} \undef\IB 
\DeclarePairedDelimiterX\IB[1]{\lbrack\!\delimsize\lbrack}{\rbrack\!\delimsize\rbrack}%
    {#1}
\reDeclarePairedDelimiterInnerWrapper\IB{nostarnonscaled}{\ifx.#1\else\mathopen#1\fi#2\ifx.#3\else\mathclose#3\fi}
\reDeclarePairedDelimiterInnerWrapper\IB{nostarscaled}{\mathopen{#1}#2\mathclose{#3}}

\newcommand\Unit {\ensuremath{\mathop{\eta}}}
\newcommand\Avg {\ensuremath{\mathop{\mu}}}

\newcommand\RealNonNeg {\ensuremath{{\mathbb R}_{+\mspace{-1mu}}}}

\newcommand\Comp {\ensuremath{\mathbin\circ}}

\newcommand\fnId {\ensuremath{\mathsf{id}}}

\newcommand\diagOf[1] {%
  \ThisStyle{\smash{\raisebox{0.0625em}{$\SavedStyle\ulcorner$}}}\mspace{-3mu}%
  #1
  \ThisStyle{\mspace{-3mu}\smash{\raisebox{-0.1875em}{$\SavedStyle\lrcorner$}}}%
}
\newcommand\fnDiag{\ensuremath{\mathsf{copy}}}

\newcommand{\fnConstZero} {\ensuremath{\mathord{\mathop{\mathbf{const}} \,\IB{\False}}}}
\newcommand{\elemZero} {\ensuremath{\mathord{\mathsf 0}}}
\newcommand{\elemOne} {\ensuremath{\mathord{\mathsf 1}}}

\DeclareMathOperator\UpperCvxCl {\mathord\uparrow\mathrm{conv}}
\DeclareMathOperator\UpperCl {\mathord\uparrow}
\newcommand\fnMap {\ensuremath{\mathsf{map}}}

\newcommand\MIN {\ensuremath{\mathbin{%
  \setlength{\fboxsep}{0pt}\raisebox{0.02em}{$\fbox{\vbox to 0.62em{\vss\hbox to 0.62em{\hss\scriptsize$\sqcap$\hss}\vss}}$}}%
}}

\newcommand\progVar[1] {\mathtt{#1}}

\newcommand\printNDivTwo{\Print 2 \lfloor \progVar{n} \!\div\! 2 \rfloor}
\newcommand\predNAndBEven{\progVar{n} \!+\! \progVar{b} \; \text{even}}
\newcommand\predNEven{\progVar{n} \; \text{even}}
\newcommand\predNOdd {\progVar{n} \; \text{odd}}

\newcommand\OP{\ensuremath{\mathord{\mathit{OP}}}}

\newcommand\fnWhileStep {\ensuremath{\mathsf{step}}}

\newcommand\ordinalFor[1] {\ensuremath{\lfloor#1\rfloor}}
\newcommand\fnOrdinalSucc {\ensuremath{\mathrm{S}}}

\DeclareMathOperator\If {\mathbf{if}}
\DeclareMathOperator\Then {\mathbf{then}}
\DeclareMathOperator\Else {\mathbf{else}}
\DeclareMathOperator\While {\mathbf{while}}
\DeclareMathOperator\Skip {\mathbf{skip}}
\DeclareMathOperator\Abort {\mathbf{abort}}
\DeclareMathOperator\HidVar {\mathbf{hid\,var}}
\DeclareMathOperator\Var {\mathbf{var}}
\DeclareMathOperator\Unvar {\mathbf{unvar}}

\DeclareMathOperator\Print {\mathbf{print}}
\DeclareMathOperator\Assert {\mathbf{assert}}

\DeclareMathOperator\Magic {\mathbf{magic}}

\newcommand\NonDet {\ensuremath{\mathbin{\sqcap}}}

\newcommand\Assign {\mathbin{\mathtt{\coloneqq}}}
\newcommand\AssignProb {\Assign}

\newcommand\Xor {\ensuremath{\mathbin{\mathbf{xor}}}}

\newcommand\rep {\ensuremath{\mathsf{rep}}}
\newcommand\repFwd {\ensuremath{\mathsf{rep}_{\mathrm{F}}}}
\newcommand\repBwd {\ensuremath{\mathsf{rep}_{\mathrm{B}}}}

\newcommand\HRef  {\ensuremath{\mathrel{\sqsubseteq}}}
\newcommand\HRefNeq   {\ensuremath{\mathrel{\sqsubsetneq}}}
\newcommand\HRefEquiv {\ensuremath{\mathrel{\equiv}}}
\newcommand\HRefR {\ensuremath{\mathrel{\sqsupseteq}}}

\newcommand\RefMDP {\ensuremath{\mathrel{\sqsubseteq_{\mathrm{MDP}}}}}
\newcommand\RefRMDP {\ensuremath{\mathrel{\sqsupseteq_{\mathrm{MDP}}}}}
\newcommand\RefEquivMDP {\ensuremath{\mathrel{\equiv_{\mathrm{MDP}}}}}
\newcommand\RefMDPNeq {\ensuremath{\mathrel{%
  \sqsubsetneq_{\mathrm{MDP}}%
}}}

\newcommand\HRefCD {\ensuremath{\HRef}}
\newcommand\HRefRCD {\ensuremath{\HRefR}}

\newcommand\DataRef {\ensuremath{\HRef}}

\newcommand\sqsubsetneq {%
  \ensuremath{\mathrel{%
  \mathclap{\vphantom{\sqsubseteq}}%
  \smash{\ThisStyle{
      \ensurestackMath{%
        \stackinset{c}{0pt}{b}{-2pt}{\scriptscriptstyle/}{%
          \SavedStyle\sqsubseteq%
        }%
      }%
  }}%
  }}%
}

\newcommand\Wpe {\ensuremath{\mathsf{wp}}}
\newcommand\WPe[1] {\ensuremath{\Wpe.#1}}
\newcommand\WPE[2] {\ensuremath{\Wpe.#1.#2}}
\newcommand\Wpl {\ensuremath{\mathsf{wpl}}}
\newcommand\WPl[1] {\ensuremath{\Wpl.#1}}
\newcommand\WPL[2] {\ensuremath{\Wpl.#1.#2}}
\newcommand\WplZ[1] {\ensuremath{\Wpl^{#1}}}
\newcommand\WPlZ[2] {\ensuremath{\WplZ{#1}\!.\mathopen{\mkern1mu}#2}}
\newcommand\WPLZ[3] {\ensuremath{\WplZ{#1}\!.\mathopen{\mkern1mu}#2.#3}}

\newcommand\Def[1] {Def.\,\ref{#1}}

\newcommand\Thm[1] {Thm.\,\ref{#1}}

\newcommand\Lem[1] {Lem.\,\ref{#1}}
\newcommand\LEM[1] {Lemma \ref{#1}}
\newcommand\Prop[1] {Prop.\,\ref{#1}}

\newcommand\COR[1] {Corollary \ref{#1}}
\newcommand\Ex[1] {Ex.\,\ref{#1}}
\newcommand\EX[1] {Example \ref{#1}}
\newcommand\Rem[1] {Rem.\,\ref{#1}}

\newcommand\Fig[1] {Fig.\,\ref{#1}}
\newcommand\FIG[1] {Figure \ref{#1}}
\newcommand\Eqn[1] {(\ref{#1})}

\newcommand\EQns[1] {Eqns.\,(\ref{#1})}
\newcommand\EQn[1] {Eqn.\,(\ref{#1})}

\newcommand\KuifjeNonDet {\ensuremath{\textrm{Kuifje}_{\NonDet}}}
\newcommand\ExpVal {\textsc{ev\@}}

\begin{document}
\title{Forward and Backward Simulations for Partially Observable Probability}
\author{Chris Chen\inst{1}\orcidID{0009-0000-2628-7343},
Annabelle McIver\inst{1}\orcidID{0000-0002-2405-9838}, \and
Carroll Morgan\inst{2}\orcidID{0000-0002-8535-9068}}
\authorrunning{C. Chen et al.}
\institute{School of Computing, Macquarie University \and
  University of New South Wales and Trustworthy Systems}
\maketitle              %
\begin{abstract}
  Data refinement is the standard
  extension of a refinement relation from programs to datatypes
  (i.e.\@ a behavioural subtyping relation).
  Forward/backward simulations
  provide a tractable method for establishing data refinement,
  and have been thoroughly studied for nondeterministic programs.
  However, for standard models of mixed probability and nondeterminism,
  ordinary assignment statements may not commute with (variable-disjoint) program fragments.
  This (1) invalidates a key assumption underlying the soundness of simulations,
  and (2) prevents modelling probabilistic datatypes with
  encapsulated state.

  We introduce a weakest precondition semantics for \KuifjeNonDet,
  a language for partially observable Markov decision processes,
  using so-called \emph{loss (function) transformers}.
  We prove soundness of forward/backward simulations in this richer setting,
  modulo healthiness conditions with a remarkable duality:
  forward simulations \emph{cannot leak} information, and
  backward simulations \emph{cannot exploit} leaked information.

\keywords{Forward and backward simulations \and Abstract datatypes
  \and Data refinement
  \and Probabilistic nondeterminism
  \and Weakest precondition
  \and Quantitative information flow
  \and Probabilistic predicates
}
\end{abstract}

\section{Introduction}
\label{sec005}

Datatype encapsulation \cite{Liskov:74,He:86ab,Hoare:87,Abadi:88} is a long-established organising principle for datatypes
that encourages defining a datatype's representation and logic
separately from a surrounding program (``program context'') $\CtxP$ that uses the datatype for its own purposes.
A \emph{data refinement} relation \cite{Roever:98}
from e.g.,
an ``abstract'' datatype $D_\AA$ to a ``concrete'' one $D_\CC$
indicates that
``the behavior of $\CtxP$ is unchanged when [$D_\CC$] is substituted for [$D_\AA$]''
\cite{Liskov:87}---%
that is,
replacing abstract with concrete preserves
all ``safety/correctness properties'' of the surrounding program
(a notion that varies with the model of computation).

A \emph{forward(/backward) simulation}
\cite{Lynch:95,He:86ab}
from $D_\AA$ to $D_\CC$
is a transformation between their internal states
that
(informally) witnesses a data refinement.
This allows a programmer
to avoid a full inductive proof
by providing a few small program refinements instead.

The probabilistic programming language pGCL \cite{Morgan:99a,McIver:05a}
combines probabilistic choice \cite{Kozen:81} with (demonic) nondeterminism.
However, in the Markov decision process (MDP) model underlying pGCL,
nondeterminism may depend on the program state.%
\footnote{This behaviour is inherited from GCL, where the absence of probability
mitigates any issues.}
This means
nondeterminism and probabilistic choice do not necessarily commute,
invalidating a key assumption underlying the soundness of forward/backward simulations \cite{Chen:24aab}.
Furthermore, MDPs are unable
to model datatypes with encapsulated state
(inaccessible to the programmer)---%
see e.g.\@ the cached random bit generator
  (\Fig{figRandBits}).

\emph{Partially observable} MDPs (POMDPs) \cite{Kaelbling:98,McIver:12}
model internal state as
\emph{hidden}
from the \NonDet-resolver.
\cite{Chen:24aab} proposed using POMDPs and techniques from
\emph{quantitative information flow} (QIF) \cite{Smith:2011ab,Alvim:20a}%
---which deals with refinement distinguishing hidden/visible state---%
as a basis
to study datatype encapsulation
for mixed probability and nondeterminism.
POMDPs provide a reasonable model for
e.g.\@ one-sided abstraction
and client-server situations---%
though this requires a
conflation of
\emph{hidden}
vs.\@
\emph{encapsulated}
(from the \NonDet-resolver vs.\@ the programmer)
which prevents full generality.
Our
aim is to provide a \underline{full theoretical basis} for \cite{Chen:24aab}.

\subsubsection{Contributions}
\label{sec018}

We provide a weakest precondition semantics
\cite{Dijkstra:76, Dijkstra:90}
for \KuifjeNonDet\@, a language for POMDPs,
using \emph{loss functions} as our predicates.
This \emph{weakest pre-loss} (\Wpl) semantics
is similar to
the leakage logic of \cite{Chen:25ab}
but additionally supports
nontermination, nondeterminism, and (type) context extension.

Our \underline{first} main results are
sound
\emph{healthiness conditions}
for our semantics
\cite{Gardiner:93,Morgan:96d}
\unskip---%
i.e.\@
equational properties
of the ``loss transformer'' functions in the image of $\Wpl$
that enable axiomatic reasoning in our proofs.
Notably, we find that
each loss transformer is superlinear, partial, and Scott-continuous
(\Lem{l101552}).
Furthermore,
under context extension,
healthy
loss transformers commute with \emph{partial} hidden assignment
on disjoint state
(\Lem{l1610}).
The former resembles
healthiness for probabilistic predicate transformers \cite[\S7]{Morgan:96d}.
The latter is a stronger version of ``correlation transformers'' \cite{Rabehaja:19ab}
that commutes with assertions.

For our \underline{second} main results
we obtain sound
forward and backward simulation rules
for \KuifjeNonDet\@
(\Thm{thm1234}, \Thm{thm1235}).
We require additional healthiness conditions on simulations, which we justify with counterexamples.
These conditions
reveal a remarkable duality:
forward simulations \emph{must not leak} information,
whereas backward simulations \emph{must not exploit} leaked information.

Section \ref{sec020} reviews data refinement for typed languages.
Section \ref{sec007} reviews MDPs and probabilistic predicate transformers,
and demonstrates how non-commutativity of probability and nondeterminism in pGCL \Eqn{e1833}
complicates data encapsulation (\Ex{ex091549}).
Section \ref{sec200} introduces \KuifjeNonDet\@ (our POMDP language),
loss functions, and our ``weakest pre-loss'' (\Wpl) semantics.
Section \ref{sec205}
gives our healthiness properties for \Wpl\@
and relates them to `hidden' and `choiceless' \KuifjeNonDet\@ programs.
Finally,
Section \ref{sec060}
gives our
forward+backward simulation rules,
and in Section \ref{sec150} we apply our results to a number of positive and negative examples
that pGCL alone cannot model.

\subsubsection{Related work}
\cite{McIver:12}
gave a
covariant ``state transformer'' semantics for POMDPs
involving powerdomains over \emph{hyperdistributions}---%
distributions of distributions.
Unfortunately,
in the hyperdistribution model
our soundness conditions
prove cumbersome to characterise and reason with;
this motivates our dual (by construction) contravariant ``weakest pre-loss'' semantics.

Much of the
early work
on simulations
for (demonic) nondeterminism
takes the relational view
\cite{He:86ab,Hoare:87,Lynch:95},
ordering programs by trace inclusion.
This is
opposite to refinement
\cite{Back:98,Dijkstra:90}.
\unskip\footnote{
  This is a ``may-must'' duality: possible executions/traces vs.\@ necessary properties.
}
A note, then, regarding nomenclature.
With respect to nondeterminism,
what we call
forward (backward) simulations
are similar to
the downward (upward) simulations of \cite{Hoare:87},
the encodings (decodings) of \cite{Back:00},
the simulations (cosimulations) of \cite{Gardiner:93},
and
the forward (backward) simulations of \cite{Lynch:95,Hasuo:06};
they
are $2$-dual to
the lax (oplax) morphisms of \cite{Hasuo:06,Johnson:09}.
\unskip\footnote{%
For partial probability our terminology disagrees with \cite{Hasuo:06},
whose orders on $\Powerset$ and $\DistPartial$
inconsistently take
\NonDet-identity ($\Magic$ \cite{Back:98}) vs.\@ divergence ($\Abort$)
as least element.
}

\section{Preliminaries}

\label{sec020}

We occasionally refer to the
Scott topology on posets,
in which open sets
are open under directed joins
(see
e.g.\@
\cite{Abramsky:94ab}).
Functions between posets
are \emph{Scott-continuous}
(``Scott-cts.'')
if they preserve (existing) directed joins;
they are necessarily monotone.

By ``state spaces'' we always mean (discrete) sets.
We
write $P \In \XX \ProgFromTo \YY$ to denote a program with
initial and final state space $\XX$ and $\YY$ respectively.

We write $\PowersetNonempty \XX$ for the nonempty powerset.
We write $\DistPartial \XX$ for the
\emph{partial probability distributions} over $\XX$
($\sum_x \delta(x) \leq 1$),
and $\Dist \XX$ for the (total) probability distributions.
Supports are countable but need not be finite.

\subsection{Refinement and nondeterminism}
\label{sec021}

A \emph{(program) refinement relation} $(\sqsubseteq)$
formalises
the notion of
behavioural subtyping
\cite{Floyd:67,Wirth:71}.
Given
compatibly-typed programs
$P, Q \In \XX \ProgFromTo \YY$ (or their denotations),
$P \sqsubseteq Q$ (``$P$ is refined by $Q$'') denotes that
replacing $P$ with $Q$ ``preserves safety/correctness properties''.
We require that $(\sqsubseteq)$ is a partial order
giving a poset-enriched category
\Eqn{e051246}.
\emph{(Demonic) nondeterministic choice} (\NonDet) is defined as the meet for this order \Eqn{e051616}.
\begin{alignat}{4}
  \label{e051246}
  P_1 &\sqsubseteq Q_1 &\;\land\; && P_2 &\sqsubseteq Q_2 &
  \;\IMP&\;
  P_1 ; P_2 \sqsubseteq Q_1 ; Q_2
  \\
  \label{e051616}
  Q &\sqsubseteq P_1 &\;\land\; && Q &\sqsubseteq P_2 &
  \;\Iff&\;
  Q \sqsubseteq P_1 \NonDet P_2
\end{alignat}
There are two common interpretations of $P_1 \NonDet P_2$.
The first, emphasising abstraction, is
\emph{programmer's choice:}
during development,
the programmer may choose between $P_1$ and $P_2$ however she pleases.
The second, emphasising small-step semantics, is \emph{runtime choice}.
In this view there is some ``\NonDet-resolver''
that chooses between $P_1$ and $P_2$ at runtime
according to some unknown strategy
which may depend on program state.
These interpretations are consistent in ordinary pGCL
but will diverge as we impose encapsulation boundaries.

\subsection{Datatypes}
\label{sec006}

Our datatypes are explicitly typed
\cite{He:86ab,Bolton:99}.
Datatype operations are interpreted
via
the copy rule \cite{Bottenbruch:62,Langmaack:80ab}
(i.e.\@ inlining);
parameters and return values are passed
using some (unencapsulated) \emph{shared state} $\SS$.
\unskip\footnote{
  Coalgebraic models \cite{Hasuo:06} might instead encode $\SS$ in the endofunctor.
}

\begin{definition}
\label{d101622}
A \underline{\emph{datatype signature}}
$(\SS, J)$ is given by a
shared state space $\SS$ and
an indexing set $J$.
\end{definition}

\begin{definition}[Datatypes \cite{He:86ab,Gardiner:93}]
\label{d8879}%
A \underline{\emph{datatype}}
with signature $(\SS, J)$
is a tuple $D = (I, \OP, F, \AA)$,
where the set $\AA$ is called the \underline{\emph{encapsulated state space}},
$I \In \SS \ProgFromTo \SS \shorttimes \AA$ and
$F \In \SS \shorttimes \AA \ProgFromTo \SS$ are programs called respectively
the \underline{\emph{initialisation}} and \underline{\emph{finalisation}},
and
$\OP = \{\OP_j \In \SS \shorttimes \AA \ProgFromTo \SS \shorttimes \AA\}_{j \In J}$
is an indexed set of programs.

Collectively, $I$, $F$, and the members of $\OP$ are called
the \underline{\emph{operations}} of the datatype.
\end{definition}

A \emph{program context} $\CtxP$ for signature $(\SS, J)$ is
``a program with $\OP_j$-shaped holes'':
a syntactic program
defined with programs
$\{\OP_j \In \SS \ProgFromTo \SS \}_{j \In J}$ in scope.
Write $\CtxP(\OP)$ for the result of inlining
$\OP$;
the types
enforce state encapsulation.

\begin{definition}[Data refinement]
\label{d8880}
    Fix signature $(\SS, J)$.
    Let $D_\AA := (I, \OP, \allowbreak F, \allowbreak \AA)$ and
    $D_\CC := (I', \allowbreak \OP', \allowbreak F', \CC)$.
    We say
    $D_\CC$ \emph{\underline{(data) refines}} $D_\AA$
    whenever
    \begin{equation}
      I ; \CtxP(\OP) ; F
      \;\sqsubseteq\;
      I' ; \CtxP(\OP') ; F'
      \label{eDataRef}
    \end{equation}
    for all program contexts $\CtxP$.
    Overloading notation, we write $D_\AA \DataRef D_\CC$.
\end{definition}

\subsection{Forward and backward simulations}

Simulations are the standard generalisation
of bisimulations from equivalence to refinement.%
\footnote{
  When
  the squares \Eqn{eqsSimCosimCD}
  are equalities
  we may speak of
  \emph{functional bisimulation} \cite{Rutten:00}.
}
Recall that a denotational semantics
maps syntactic programs to
a category of so-called \emph{abstract programs} (e.g.\@ predicate transformers).
A forward or backward simulation
is then an
abstract program
as follows:

\begin{definition}
\label{d071245}
Let $D_\AA = (I, \OP, F, \AA)$ and $D_\CC = (I', \OP', F', \CC)$
be datatypes with shared signature $(\SS, J)$.
A \underline{\emph{forward simulation}} from $D_\AA$ to $D_\CC$
is
an abstract program $\repFwd \In \AA \ProgFromTo \CC$
such that the refinements \Eqn{eSimCD} hold:
\expandafter\xlabel[\ref*{eSimCD}--\ref*{eCosimCD}]{eqsSimCosimCD}%
\begin{equation}
\label{eSimCD}
\begin{tikzcd}[
  refinementlabel/.style={scale=1},
  column sep=small,
  row sep=1.5em,
]
  \SS
    \arrow[r, "I"]
    \arrow[dr, "I'" swap, out=-90, in=180]
    \arrow[d,phantom,"" {name="rep0"}]
  &
  \SS \cramped\times \AA
  \arrow[d, "\repFwd"]
    \arrow[d,phantom,"" {name="rep1"}]
  \\
  {}
  &
  \SS \cramped\times \CC
  \arrow[phantom, from="rep0", to="rep1", "\HRefRCD" refinementlabel]
\end{tikzcd}
\qquad
\begin{tikzcd}[
  refinementlabel/.style={scale=1},
  column sep=1.8em,
  row sep=1.5em,
]
  \SS \cramped\times \AA \arrow[d, "\repFwd" swap] \arrow[r, "OP_j"]
    \arrow[d,phantom,"" {name="rep0"}]
  &
  \SS \cramped\times \AA
  \arrow[d, "\repFwd"]
    \arrow[d,phantom,"" {name="rep1"}]
  \\
  \SS \cramped\times \CC \arrow[r, "OP'_j" swap]
  &
  \SS \cramped\times \CC
  \arrow[phantom, from="rep0", to="rep1", "\HRefRCD" refinementlabel]
\end{tikzcd}
\qquad
\begin{tikzcd}[
  refinementlabel/.style={scale=1},
  column sep=small,
  row sep=1.5em,
]
  \SS \cramped\times \AA \arrow[d, "\repFwd" swap]
    \arrow[dr, "F", out=0, in=90]
    \arrow[d,phantom,"" {name="rep0"}]
  &
  {}
    \arrow[d,phantom,"" {name="rep1"}]
  \\
  \SS \cramped\times \CC \arrow[r, "F'" swap]
  &
  \SS
  \arrow[phantom, from="rep0", to="rep1", "\HRefRCD" refinementlabel]
\end{tikzcd}
\end{equation}
and a \underline{\emph{backward simulation}} from $D_\AA$ to $D_\CC$
is
an abstract program $\repBwd \In \CC \ProgFromTo \AA$
such that the refinements \Eqn{eCosimCD} hold.
\begin{equation}
\label{eCosimCD}
\begin{tikzcd}[
  refinementlabel/.style={scale=1},
  column sep=small,
  row sep=1.5em,
]
  {}
    \arrow[d,phantom,"" {name="rep0"}]
  &
  \SS \cramped\times \AA
    \arrow[d,phantom,"" {name="rep1"}]
  \\
  \SS \arrow[r, "I'" swap]
    \arrow[ur, "I", out=90, in=180]
  &
  \SS \cramped\times \CC \arrow[u, "\repBwd" swap]
  \arrow[phantom, from="rep0", to="rep1", "\HRefCD" refinementlabel]
\end{tikzcd}
\qquad
\begin{tikzcd}[
  refinementlabel/.style={scale=1},
  column sep=1.8em,
  row sep=1.5em,
]
  \SS \cramped\times \AA \arrow[r, "OP_j"]
    \arrow[d,phantom,"" {name="rep0"}]
  &
  \SS \cramped\times \AA
    \arrow[d,phantom,"" {name="rep1"}]
  \\
  \SS \cramped\times \CC \arrow[r, "OP'_j" swap]
  \arrow[u, "\repBwd"]
  &
  \SS \cramped\times \CC \arrow[u, "\repBwd" swap]
  \arrow[phantom, from="rep0", to="rep1", "\HRefCD" refinementlabel]
\end{tikzcd}
\qquad
\begin{tikzcd}[
  refinementlabel/.style={scale=1},
  column sep=small,
  row sep=1.5em,
]
  \SS \cramped\times \AA \arrow[r, "F"]
    \arrow[d,phantom,"" {name="rep0"}]
  &
  \SS
    \arrow[d,phantom,"" {name="rep1"}]
  \\
  \SS \cramped\times \CC \arrow[ur, "F'" swap, out=0, in=-90]
  \arrow[u, "\repBwd"]
  &
  {}
  \arrow[phantom, from="rep0", to="rep1", "\HRefCD" refinementlabel]
\end{tikzcd}
\end{equation}
\end{definition}
Note that, according to the types,
$\rep$ can \emph{only} act on encapsulated state,
not on shared state.
The $\rep$ arrows above
represent (type-)context extensions of $\rep$ to a state space including $\progVar{s} \In \SS$.

As \EQns{eqsSimCosimCD} hint,
soundness of simulation
is often shown by induction over $\CtxP$.
This requires a property for interleaving independent programs
$P \In \XX \ProgFromTo \YY$:
\begin{equation}
  \label{eSimCosimWeakCommut}
\begin{tikzcd}[
  refinementlabel/.style={scale=1},
  column sep=1.8em,
  row sep=1.5em,
]
  \XX \cramped\times \AA \arrow[d, "\repFwd" swap] \arrow[r, "P"]
    \arrow[d,phantom,"" {name="rep0"}]
  &
  \YY \cramped\times \AA
  \arrow[d, "\repFwd"]
    \arrow[d,phantom,"" {name="rep1"}]
  \\
  \XX \cramped\times \CC \arrow[r, "P" swap]
  &
  \YY \cramped\times \CC
  \arrow[phantom, from="rep0", to="rep1", "\sqsupseteq" refinementlabel]
\end{tikzcd}
\quad
\begin{tikzcd}[
  refinementlabel/.style={scale=1},
  column sep=1.8em,
  row sep=1.5em,
]
  \XX \cramped\times \AA \arrow[r, "P"]
    \arrow[d,phantom,"" {name="rep0"}]
  &
  \YY \cramped\times \AA
    \arrow[d,phantom,"" {name="rep1"}]
  \\
  \XX \cramped\times \CC \arrow[r, "P" swap]
  \arrow[u, "\repBwd"]
  &
  \YY \cramped\times \CC \arrow[u, "\repBwd" swap]
  \arrow[phantom, from="rep0", to="rep1", "\HRefCD" refinementlabel]
\end{tikzcd}
\end{equation}

\begin{example}
  \label{ex1508}
  It is well-known that
  for guarded nondeterminism,
  e.g.\@
  $\Kl(\Powerset)$ \cite{Lynch:95} and $\Kl(\Powerset + \bot)$ \cite{Hoare:87,He:86ab},
  \Eqn{eSimCosimWeakCommut} holds.
  In these settings,
  certain forward and backward simulations
  are sound, and
  composition of backward-then-forward simulations is complete.
  For $\Kl(\PowersetNonempty + \bot)$,
  under
  reasonable assumptions about finalisations,
  backward simulations are complete
  \cite{Gardiner:93}.
\end{example}

\section{MDPs}
\label{sec007}

We briefly review pGCL, a language for
Markov decision processes (MDPs)
mixing nondeterminism and probability \cite{Morgan:99a,McIver:05a}.
Assignment ($\AssignProb$) takes probabilistic expressions on the right.
As syntactic sugar we write nondeterministic choice between \emph{expressions} (instead of programs),
e.g.\@ ``$\progVar{x} \Assign 0 \NonDet 1$''
is \emph{shorthand} for $(\progVar{x} \Assign 0) \NonDet (\progVar{x} \Assign 1)$.

Per ``programmer's choice'',
the \NonDet-resolver may employ a ``mixed'' (probabilistic) strategy \Eqn{e1347b},
or depend
on the state \Eqn{e1347a}.
(Below, $\progVar{b} \In \{0, 1\}$,
and
pGCL refinement (\RefMDP) and equivalence (\RefEquivMDP) are to be defined.)
\begin{gather}
  \label{e1347b}
  \progVar{x} \Assign 0 \NonDet 1
  \quad\RefMDP\quad \progVar{x} \Assign 0 \PC{\frac{2}{3}} 1
  \\
  \label{e1347a}
  \progVar{x} \Assign 0 \NonDet 1
  \quad\RefEquivMDP\quad \progVar{x} \Assign \progVar{b} \NonDet \lnot\progVar{b}
  \quad\RefMDP\quad \progVar{x} \Assign \progVar{b}
\end{gather}

Operationally, pGCL programs of type $\XX \ProgFromTo \YY$ are modellable as
$\XX \Fun \PowersetNonempty \DistPartial \YY$ functions,
normalised by upper convex closure.

\subsection{Probabilistic predicates and pGCL semantics}
\label{sec070}

\subsubsection{Preliminaries}
The \emph{extended (non-negative) reals} are $\RExtd := [0, \infty]$.
They form a semiring (``rig'') under the usual addition/multiplication,
with all operations monotone (indeed, Scott-cts.\@).
Recall a \emph{d-cone} \cite{Tix:09ab} is
a directed-complete `$\RExtd$-semimodule'
$(C, \leq, +, \elemZero, \cdot)$
equipped with
Scott-cts.\@
$\RExtd$-scalar multiplication
and
a Scott-cts.\@ addition monoid,
that obey various distributivity laws.

\smallskip

The \emph{probabilistic predicates} over state space $\XX$ are
the ``extended random variables''
$\Pred \XX := (\XX \Fun \RExtd)$.
They form a d-cone with all operations pointwise.
We write $\fnId_\XX$ for the identity function on $\Pred \XX$.

Given a Boolean expression $\BB$ with
$\progVar{x} \In \XX$ free,
the \emph{indicator function}
$\IB{\BB}_\XX \In \Pred \XX$
takes $x \mapsto 1$ if $\BB[\progVar{x}/x]$ is true, and
$x \mapsto 0$ otherwise.
(We generally omit the subscript.)
As a special case,
let
$\elemOne_\XX = \IB{\True}_\XX$
and
$\elemOne_x = \IB{\progVar{x} = x}_\XX$.
When $e \leq \elemOne_\XX$, write $\lnot e$ for the unique solution to $e + \lnot e = \elemOne_\XX$.
The \emph{conjunction} monoid $(\Pred \XX, \boxtimes, \elemOne_\XX)$
is given by
$(e_1 \boxtimes e_2)(x) = e_1(x) e_2(x)$.

Given $e \In \Pred \XX$,
its \emph{context extension}
$e^\ZZ \In \Pred (\XX \shorttimes \ZZ)$
is $e^\ZZ(x, z) = e(x)$.
\label{paraTensor}
Given linear, Scott-cts.\@
functions
$f \In \Pred \ZZ \Fun \Pred \XX, f' \In \Pred \WW \Fun \Pred \YY$,
their \emph{tensor}
$f \otimes f' \In \Pred (\ZZ \shorttimes \WW) \Fun \Pred (\XX \shorttimes \YY)$
is the unique linear, Scott-cts.\@
function s.t.\@
$(f \otimes f')(\elemOne_{(z,w)})(x, y) = f(\elemOne_z)(x) \times f'(\elemOne_w)(y)$.

\smallskip
Every function $f \In \XX \Fun \DistPartial \YY$
has a
\emph{dual}
$f^\intercal \In \Pred \YY \Fun \Pred \XX$
given by
$f^\intercal(e)(x) = \sum_{y} e(y) \times f(x)(y)$.
The dual
is linear, Scott-cts.\@,
and \emph{partial}
($f^\intercal(\elemOne_\YY) \leq \elemOne_\XX$);
if $f \In \XX \Fun \Dist \YY$
is \emph{total},
then so is its dual
($f^\intercal(\elemOne_\YY) = \elemOne_\XX$).

Let $\fnDiag_\YY \In \YY \Fun \Dist \YY^2$ be zero everywhere except $\fnDiag_\YY(y)(y,y) = 1$.

Given a weight $p \in [0, 1]$,
we write $\delta \PC{p} \delta'$ for the ``convex combination''
of two compatible probability distributions,
$(\delta \PC{p} \delta')(x) = p \delta(x) + (1-p) \delta'(x)$.

\subsubsection{Weakest preexpectation (\Wpe)}

Just as Dijkstra gave the semantics of GCL
in terms of weakest precondition Boolean ``predicate transformers'' \cite{Dijkstra:76,Dijkstra:90},
the semantics of pGCL can be given in terms of \emph{probabilistic predicate transformers}
\cite{Morgan:96d,Jones:89ab}
\unskip---%
for our purposes, with mutable state \cite{Ye:21}.

\label{para1813}
For each pGCL program $P \In \XX \ProgFromTo \YY$
there is a \emph{weakest preexpectation} function $\WPe{P} \In \Pred \YY \Fun \Pred \XX$
\cite[\S4, Fig.\@ 2]{Morgan:96d}.
The \emph{refinement relation} $(\RefMDP)$ over two pGCL programs
is defined as $P \RefMDP Q$ whenever $\WPe{P} \leq \WPe{Q}$ pointwise.

\begin{remark}[Interpreting \Wpe]
  \label{r051116}
  Informally,
  the expected value (``\ExpVal'') of
  $e$
  after executing $P$
  is $\geq r$ for all \NonDet-resolvers
  \emph{iff}
  the initial
  \ExpVal\@ of $\WPE{P}{e}$
  is $\geq r$.

  If $P$ doesn't terminate,
  we say the \ExpVal\@ of $e$ is zero.
  Hence $\WPE{\Abort}{e} = \elemZero_\XX$,
  and
  $\WPE{P}{\elemOne_\YY}$
  is the vector of \emph{termination probabilities} $\Pr(\Terminates \!\mid\! \progVar{x} = x)$.
\end{remark}

\begin{example}
  \label{ex1713}
  Given
  $r \In \RExtd$, $e_\YY \In \Pred \YY$,
  $\BB \In \XX \Fun \Bool$,
  we have $r \IB{\BB} \leq \WPE{P}{e_\YY}$ iff the
  final
  \ExpVal\@ of $e_\YY$ is at least $r$ whenever $\BB$ initially holds.
\end{example}

\subsection{pGCL, hidden state, and noncommutativity}
\label{sec071}

pGCL
has only \emph{visible} state:
the nondeterminism may always depend directly on the state.
We have the following \emph{strict} inequality:
\begin{gather}
  (
  \progVar{y} \AssignProb 0 \PC{p} 1;
  \;
  \progVar{x} \Assign 0 \sqcap 1
  )
\quad
\RefMDPNeq
\quad
  (
  \progVar{x} \Assign 0 \sqcap 1;
  \;
  \progVar{y} \AssignProb 0 \PC{p} 1
  )
\label{e1833}
\end{gather}

Since
probabilistic and nondeterministic choice do not commute,
\EQn{eSimCosimWeakCommut} doesn't hold in general:
simulations are not sound for pGCL.

\begin{example}[Strictness of \Eqn{e1833}]
\label{ex091759}
Write $P$ and $Q$ for the left/right programs of \Eqn{e1833} respectively.
After running $Q$,
we are guaranteed that $\Pr(\progVar{x} = \progVar{y}) = \frac{1}{2}$
(as any \NonDet-resolver for $Q$ is independent of
the final value of $\progVar{y}$).
However, for the left program
we cannot guarantee that $\Pr(\progVar{x} = \progVar{y}) > 0$
as a \NonDet-resolver
may potentially always choose $\progVar{x} \Assign \lnot \progVar{y}$.
Then
$\WPE{P}{\IB{\progVar{x} = \progVar{y}}} \not \geq \WPE{Q}{\IB{\progVar{x} = \progVar{y}}}$,
and therefore
$P \not \RefRMDP Q$.

On the other hand,
a simple strategy stealing argument between \NonDet-resolvers
gives $P \RefMDP Q$.
\end{example}

\begin{remark}
\label{r091344}
  Strictness
  of \Eqn{e1833}
  reduces algebraically to the asymmetry in pGCL of
  \begin{equation}
    \label{e1834}
    P ; (Q \NonDet R) \RefMDP (P ; Q) \NonDet (P ; R)
  \end{equation}
  when $P$ contains probabilistic choice.
  In other words, the problem is that $(\mathord;)$ fails to left-distribute over $(\NonDet)$.
  This amounts to the well-known nonexistence
  of a distributive law $\Dist \Powerset \Fun \Powerset \Dist$.
  \unskip\footnote{
  We thank the anonymous referee \#3 for pointing out this connection.}
  \unskip\footnote{
  Indeed, it turns out
  the $(\not \RefRMDP)$ argument in \Ex{ex091759}
  is isomorphic to
  the nonexistence proof
  given in \citeauthor{Varacca:03}'s thesis \cite{Varacca:03}.
  (There the construction corresponding to the left of $\Eqn{e1833}$
    is attributed to Plotkin.)
  }
\end{remark}

\smallskip
Visible state also poses a problem for modelling forms of encapsulation
which \emph{ought} to hide information from the resolver.

\begin{example}
\label{ex091549}
In \Fig{figRandBits},
the datatype $D_\AA := (I_\AA, \OP_\AA, F_\AA, \{\ast\})$
specifies a random bit generator ($\{\ast\}$ being the trivial 1-element state space),
while
$D_\CC := (I_\CC, \OP_\CC, F_\CC, \Bool)$
is a potential implementation that precomputes random bits ahead of time.
We wish to model
``programmer's choice'' with encapsulated state,
so the cached bit $\progVar{b}$ should \emph{hidden} from the \NonDet-resolver
such that the precomputation has no effect on functionality,
i.e.\@ $D_\AA \sqsubseteq D_\CC$.

However, pGCL does not provide a way to model $b$ as hidden.
Let
$\CtxP(\OP) = (\progVar{x} \Assign 0 \sqcap 1; \OP; \progVar{y} \Assign \progVar{s})$.
The same argument as for \EQn{e1833}
shows that $I_\AA; \CtxP(\OP_\AA); \allowbreak F_\AA \allowbreak \not \sqsubseteq I_\CC; \CtxP(\OP_\CC); F_\CC$.
Hence pGCL alone is insufficient to model datatypes with hidden precomputation.
\end{example}

\begin{figure}[t]%
\begin{gather*}
\text{Shared state:} \quad
\progVar{s} \In \{0,1\}.
\\
\setlength\arraycolsep{.5em}
\begin{array}{c|c}
  D_\AA & D_\CC \\ \hline
\begin{aligned}[t]
  I_\AA &\colon \mathopen{}\Skip; \\[0.33em]
  \OP_\AA &\colon \progVar{s} \Assign 0 \PC{\frac{1}{2}} 1; \\[0.33em]
  F_\AA &\colon \mathopen{}\Skip;
\end{aligned}
        &
\begin{aligned}[t]
  I_\CC &\colon
    \mathopen{}
    \Var \progVar{b} \Assign 0 \PC{\frac{1}{2}} 1 ; \\[0.33em]
  \OP_\CC &\colon
    \begin{lgathered}[t]
    \progVar{s} \Assign \progVar{b} ; \\
    \progVar{b} \Assign 0 \PC{\frac{1}{2}} 1;
    \end{lgathered} \\[0.33em]
  F_\CC &\colon \mathopen{} \Unvar \progVar{b};
\end{aligned}
\end{array}
\end{gather*}
\caption[(PLACEHOLDER for \textbackslash listoffigures]{%
  An abstract ``random bit generator''
  $D_\AA$, and
  a concrete implementation
  $D_\CC$
  precomputed with hidden variables.
  Under the copy rule,
  $D_\CC$'s encapsulated state
  is visible in pGCL,
  but can be modelled as hidden in \KuifjeNonDet\@ (\Ex{ex1538}).
  \label{figRandBits}
}
\end{figure}

\section{POMDPs}
\label{sec200}

Ideally, we want to model
a mix of visible and \emph{hidden} variables in our datatypes.
(When a variable is \emph{hidden} then the \NonDet-resolver can at best infer its value based
on probabilistic inference and side channel information.)
\unskip\footnote{
Although we conflate them in our
datatypes
and
examples (\S\ref{sec060}--\ref{sec150}),
the
\emph{visible vs.\@ hidden} distinction
is orthogonal to
\emph{global vs.\@ encapsulated}.
The former relates to
runtime choice
(the \NonDet-resolver's knowledge);
the latter, to programmer's choice.
}

\emph{Partially observable MDPs} (POMDPs)
are traditionally MDPs that
probabilistically emit \emph{observations} with every state transition \cite{Kaelbling:98};
the \NonDet-resolver
only knows the observation history, not the underlying state.
We use POMDPs to model mixed
nondeterminism and probability
with partially visible state.

Hereon we interpret nondeterminism as \emph{runtime choice}:
the \NonDet-resolver may condition on the ``program trace'' in a way
a programmer cannot
(except at the expense of modularity and encapsulation).

\begin{figure}[pt]
  \begin{gather*}
    \label{eTypeSkipAbortComp}
    \dfrac{}{\Skip \In \XX \ProgFromTo \XX}
    \quad
    \dfrac{}{\Abort \In \XX \ProgFromTo \YY}
    \quad
    \dfrac{P \In \XX \ProgFromTo \YY \quad Q \In \YY \ProgFromTo \ZZ}
      {P ; Q \In \XX \ProgFromTo \ZZ}
    \\[.5em]
    \label{eTypeAssignVarUnvar}
    \dfrac{f \In \XX \Fun \DistPartial \XX}{\progVar{x} \AssignProb f(\progVar{x}) \In \XX \ProgFromTo \XX}
    \quad
    \dfrac{f \In \YY \Fun \DistPartial \XX}{\HidVar \progVar{x} \AssignProb f(\progVar{y}) \In \YY \ProgFromTo \YY \!\times\! \XX}
    \quad
    \dfrac{}{\Unvar \progVar{x} \In \YY \!\times\! \XX \ProgFromTo \YY}
    \\[.5em]
    \label{eTypeNondetIf}
    \dfrac{P_1 \In \XX \ProgFromTo \YY \quad P_2 \In \XX \ProgFromTo \YY}
      {P_1 \NonDet P_2 \In \XX \ProgFromTo \YY}
    \quad
    \dfrac{g \In \Pred \XX \land g \leq \elemOne_\XX
      \quad P \In \XX \ProgFromTo \YY
      \quad Q \In \XX \ProgFromTo \YY
    }
      {\If g(\progVar{x}) \Then P \Else Q \In \XX \ProgFromTo \YY}
    \\[.5em]
    \label{eTypeWhilePrint}
    \dfrac{g \In \Pred \XX \land g \leq \elemOne_{\XX} \quad P \In \XX \ProgFromTo \XX}
      {\While g(\progVar{x}) \, \{ P \} \In \XX \ProgFromTo \XX}
    \quad
    \dfrac{f \In \XX \Fun \DistPartial \Obsvs
    }
      {\Print f(\progVar{x}) \In \XX \ProgFromTo \XX}
  \end{gather*}
  \caption[IGNOREME]{%
    Simplified program syntax for \KuifjeNonDet.
    We elide (standard) type contexts:
    it is assumed that $\progVar{x} \In \XX, \progVar{y} \In \YY$, etc.,
    and that $\progVar{x} \In \XX$ may stand for multiple variables
    $(\progVar{x}_i)_{i \in I} \In \prod_{i} \XX_i$.
    We equivocate betwen functions $f(\progVar{x})$ and expressions with $\progVar{x} \In \XX$ free.
    \unskip\label{figSyntax}
  }
\end{figure}

\smallskip

``\KuifjeNonDet'',
our language for POMDPs,
extends the Kuifje language for HMMs \cites{FoPPS:19}[\S8.3--4]{McIver:2014ab}[\S14.4]{Alvim:20a}
with
\begin{enumerate*}[(i),font=\itshape,ref=\emph{(\roman*)}]
  \item \label{i1305b}
    nondeterminism,
  \item \label{i1305a}
    explicit syntax for
    declaring and freeing variables, and
  \item \label{i1305c}
    context extension.
\end{enumerate*}
\KuifjeNonDet\@ looks similar to, but is not an extension of, pGCL:
all variables are intrinsicially \emph{hidden}.

Operationally,
\KuifjeNonDet\@ programs $P \In \XX \ProgFromTo \YY$
may be considered
$\Dist \XX \Fun \PowersetNonempty \DistPartial \Dist \YY$ functions
\cite{McIver:12,McIver:2014ab}.
Loosely speaking,
$P$ carries
an implicit \emph{observation space} $\Obsvs$
(in the automata model, the observable traces).
The worst-case \NonDet-resolver
begins with prior belief $\delta \In \Dist \XX$.
Then, provided that $P$ terminates,
it performs Bayesian inference,
combining the observation $\omega \in \Obsvs$ (the outer $\DistPartial$)
with knowledge of source code
to obtain a posterior belief (the inner $\Dist \YY$).
(See also \Rem{r091804}.)

\smallskip

\FIG{figSyntax} gives our syntax.
Per convention, $\Skip \In \XX \ProgFromTo \XX$ is the identity for composition,
and $\Abort \In \XX \ProgFromTo \YY$ is the nonterminating program, alt.\@ ``undefined behaviour''
(hence the bottom element for refinement).
Sequential composition $(\mathord ;)$
and nondeterminism $(\NonDet)$ are standard.

As with pGCL,
assignment is probabilistic; e.g.\@
$(\progVar{x} \AssignProb \progVar{y}\cramped+1 \PC{\frac{4}{5}} \progVar{z})$
sets $\progVar{x}$ to $\progVar{z}$ with $\tfrac{1}{5}$ probability.
If $f$ is not total ($f^\intercal \elemOne_\YY \neq \elemOne_\XX$)
then $\progVar{x} \Assign f(\progVar{x})$ `$\Abort$'s with the remaining probability.
$\HidVar$ and $\Unvar$ are variations
mutating the state space (/type context).

The $\Print$ instruction makes information available to the $\NonDet$-resolver.
Control flow
($\If$ and $\While$) is also considered \emph{visible}:
only thus can we avoid merging
observation types between different branches
(a failure of modularity).

As syntactic sugar, we introduce an $\Assert$ statement \Eqn{e101104b}.
Here, given $g \leq \elemOne_\XX$
we define $\diagOf{g} \In \XX \Fun \DistPartial \XX$
as $\diagOf{g}(x)(x) = g(x)$ and $x \neq x' \Imp \diagOf{g}(x)(x') = 0$.
\begin{align}
  \label{e101104b}
  \Assert g(\progVar{x}) \;&=\;
  (\progVar{x} \AssignProb \diagOf{g}(\progVar{x}))
\end{align}

\begin{example}
\label{ex1730}
In $\KuifjeNonDet$,
\emph{``it is safer to reveal nothing than something''}
\Eqn{e1730a},
\emph{``$\Print$ is seldom idempotent''} \Eqn{e1730b},
and isomorphic but syntactically distinct leaks (e.g.\@ of a variable $\progVar{b} \In \{0,1\}$)
are equivalent \Eqn{e1730c}.
\begin{align}
  \label{e1730a}
  \Print f(\progVar{x}) &\sqsubseteq \Skip \\
  \label{e1730b}
  \Print f(\progVar{x}) ; \Print f(\progVar{x}) &\sqsubseteq \Print f(\progVar{x})
  \\
  \label{e1730c}
  \Print \progVar{b} &\HRefEquiv \Print \lnot \progVar{b}
\end{align}
When $f$ is \emph{total} (terminates with $\Pr = 1$),
inequality \Eqn{e1730a} is strict unless $f$ is a constant function,
and \Eqn{e1730b} is strict unless $f$ is deterministic.

We remark that the refinements above
compare programs with incompatible ``observation spaces'':
the observations are not part of the type signature.
\end{example}

\begin{remark}
  \label{r101101}
  \KuifjeNonDet\@ is not quite a conservative extension of pGCL.
  A pGCL program
  can be transformed into
  a $\KuifjeNonDet$ program
  by $\Print$-ing the entire state before and after every statement.
  This transformation is sound (reflects refinement)
  but is either incomplete
  (failing to preserve the refinement $\Skip \RefMDP \progVar{v} \AssignProb \progVar{v}$)
  or non-functorial (does not preserve $\Skip$).
\end{remark}

\subsection{Loss functions}
\label{sec210}

What is the analogue of the probabilistic predicates $\Pred \XX$ for POMDPs?

\begin{example}
\label{ex052207}
Let
$P \In \Ints_4 \ProgFromTo \Ints_4 \times \Ints_2
:= \bigl(
  \printNDivTwo; \allowbreak
  \Var \progVar{b} \AssignProb 0 \NonDet 1
\bigr)$.
The $\NonDet$-resolver
chooses $\progVar{b}$
depending on the observed higher-order bit of $\progVar{n}$:
write
$f \In \{0, 2\} \Fun \Dist \Ints_2$
for its (unknown) strategy.

Following \Rem{r051116}
we consider necessary and sufficient conditions for
$\Pr(\predNAndBEven) \geq r$.
Let $\delta \In \Dist \Ints_4$
be the initial distribution of $\progVar{n}$.
Suppose $\delta(0) \leq \delta(1)$
and $\delta(3) \leq \delta(2)$.
If the \NonDet-resolver chooses $f(0) = \Unit 0$ (the point distribution on $0$)
and $f(2) = \Unit 1$,
this minimises $\Pr(\predNAndBEven)$.
Hence if $\Pr(\progVar{n} = 0) + \Pr(\progVar{n} = 3) < r$,
then in the worst case $\Pr(\predNAndBEven) < r$ also.

\begingroup
\clubpenalty=-100
\widowpenalty=-200
By similar reasoning,
$\Pr(\predNAndBEven) \geq r$ for all resolver strategies $f$
if and only if \emph{four} `pre-'predicates
have \ExpVal\@ $\geq r$: these are
$\IB{\progVar{n} = 0} + \IB{\progVar{n} = 2}$,
$\IB{\progVar{n} = 0} + \IB{\progVar{n} = 3}$,
$\IB{\progVar{n} = 1} + \IB{\progVar{n} = 2}$, and
$\IB{\progVar{n} = 1} + \IB{\progVar{n} = 3}$.

\endgroup

\end{example}

\EX{ex052207} motivates
using (nonempty) \emph{sets} of predicates $E \in \PowersetNonempty \Pred \XX$
for \KuifjeNonDet's ``predicate transformer'' semantics.
By analogy with \Rem{r051116} we evaluate these against priors/posteriors
via judgements
``$r \leq \inf_{e \in E} (\text{\ExpVal\@ of $e$})$''.
But since we \emph{only} evaluate such sets by taking minimum $\ExpVal$s,
there is some redundancy here.
(E.g.\@ $\{\IB{\progVar{x} \cramped= 1}, \IB{\progVar{x} \cramped= 2}\}$
and
$\{
  \IB{\progVar{x} \cramped= 1}, \IB{\progVar{x} \cramped= 2}, \,\cramped{\frac{1}{2}\IB{\progVar{x} = 1} + \frac{1}{2} \IB{\progVar{x} \cramped= 2}}, \allowbreak \IB{\progVar{x} \geq 2}
\}$
have the same minimum $\ExpVal$ against all distributions.)
An appropriate \emph{normal form} for $\PowersetNonempty \Pred \XX$
would give a partial order based on min.\@ \ExpVal\@s,
allowing
e.g.\@ least fixed points for $\While$ (alongside other tools from domain theory).

Algebraically, we want to
equip $\Pred \XX$ with a \emph{formal meet} ($\MIN$)
that is indifferent to convex combinations,
while preserving d-cone structure.
The ``Smyth powercone'' \cite[Thm.\@ 4.23]{Tix:09ab} provides this construction,
which we use to define our (normalised) \emph{loss functions} (\Def{d071224}).

\begin{definition}[Loss functions]
\label{d071224}
The ``loss functions'' over $\XX$, written $\Loss \XX$,
are the
nonempty, Scott-compact, convex, upper subsets of $\Pred \XX$.
Their canonical refinement order $(\sqsubseteq)$ is given by \underline{reverse} set inclusion.
\end{definition}

\begin{remark}
  \label{r091804}
  These are named after
  the loss functions of Bayesian decision theory
  \cite{Zellner:86},
  which
  \cite{Smith:2009ab,Alvim:2012aa}
  first applied to QIF.
  Given $E \In \Loss \XX$,
  the score of an (optimal) Bayes estimator
  for the Bayesian loss function with actions $E$
  is $\inf_{e \In E} (\text{\ExpVal\@ of $e$})$.
  We interpret the worst-case $\NonDet$-resolver
  as such an estimator.

  This gives a bijection of $\Loss \XX$ with
  the superlinear Scott-cts.\@ functionals
  $\DistPartial \XX \Fun \RExtd$
  (see e.g.\@ \cite{Tix:09ab}).%
  \footnote{
    Taking Bayes risk
    for a ``hyperdistribution'' $\DistPartial \Dist \XX$
    gives a unique linear Scott-cts.\@
    $U \In \DistPartial \Dist \XX \Fun \RExtd$
    satisfying
    $\forall \nabla \In \DistPartial \Dist^2 \XX \,\cdot\,
    U(\Avg \!\nabla) \leq U(\DistPartial\!\Avg \!\nabla)$:
    this is
    the \emph{data processing inequality}
    \cite{McIver:12,Alvim:20a}.
    \cite[\S4.4]{Perrone:18} studies duality for such ``affine extensions''.
  }
  Refinement is equivalent to the pointwise ordering,
  and addition and infima can be interpreted pointwise on these functionals.
\end{remark}

\begin{remark}
  \label{r061723}
The predicates of \cite{Chen:25ab}
are \emph{gain} functions
(evaluated via \emph{max} \ExpVal).
Under their model nontermination is angelic: e.g.\@
``$\Print \progVar{b} \sqsubseteq \Skip \sqsubseteq \Assert \progVar{b}$''
whereas ``$\Assert$'' should be least.
Using loss functions resolves this,
and our semantics agrees with existing
accounts of nontermination in QIF \cite[Def.\@ 7.4]{McIver:2014ab}.
\end{remark}

We have a linear Scott-cts.\@ injection of $\Pred \XX \hookrightarrow \Loss \XX$
given by principal filters: $e \mapsto \UpperCl \{e\}$,
which we treat as an \emph{embedding},
e.g.\@ writing $\IB{\BB}$ as shorthand for $\UpperCl \{\IB{\BB}\}$.
The embedding does not preserve meets; to avoid ambiguity we write meets of loss functions as $\MIN$.

Given a nonempty $E \subseteq \Pred \XX$, write $\UpperCvxCl E \in \Loss \XX$
for its upper convex closure.
The d-cone structure of $\Loss \XX$ and its finite meets
are as follows
\cite{Tix:09ab}:
\begin{align}
  \label{e1731a}
  r \cdot E_1 + E_2 &= \UpperCvxCl \setcmp{r \cdot e_1 + e_2}{e_1 \in E_1, e_2 \in E_2}
  \\
  \label{e1731e}
  E_1 \MIN E_2 &= \UpperCvxCl (E_1 \cup E_2)
  \\
  \label{e1731c}
  {\textstyle \bigvee^\uparrow_{i \in I}} E_i &= \mathop{\textstyle\bigcap}\nolimits^\downarrow_{i \in I} E_i
\end{align}
Given a linear, Scott-cts.\@ probabilistic predicate transformer
$f \In \Pred \YY \Fun \Pred \XX$,
it extends to a loss transformer
$\fnMap(f) \In \Loss \YY \Fun \Loss \XX$,
with
$
\fnMap(f)(E) = \UpperCl f(E)
$.

\begin{proposition}[{\cite[Prop.\@ 4.19]{Tix:09ab}}]
\label{p101646}
  Suppose $f \In \Pred \YY \rightarrow \Pred \XX$ is
  linear and Scott-cts.
  Then
  $\fnMap(f)$
  is
  linear,
  Scott-cts.\@,
  and
  \underline{\emph{\MIN-preserving}},
  i.e.\@
  $\fnMap(f) \allowbreak (E_1 \MIN E_2) = \fnMap(f)(E_1) \MIN \fnMap(f)(E_2)$.
\end{proposition}

Following \cite{Chen:25ab} we also extend the \emph{conjunction} operator
$\boxtimes$ pointwise to an action on $\Loss \XX$:
let $e \boxtimes E = \fnMap(e \boxtimes -)(E)$.

\subsection{Weakest pre-loss semantics}
\label{sec220}

\Def{d102024} gives
a loss transformer semantics
in terms of a family of ``weakest pre-loss''
functions $\WPlZ{\ZZ}{P} \In \Loss (\YY \shorttimes \ZZ) \Fun \Loss (\XX \shorttimes \ZZ)$.
The parameter $\ZZ$ provides the \emph{context extension}
needed for e.g.\@ evaluating $\CtxP(\OP)$.

\begin{definition}[Weakest pre-loss]
  \label{d102024}
  The \underline{\emph{weakest pre-loss}}
  for $\KuifjeNonDet$ program $P \In \XX \ProgFromTo \YY$
  and correlated state space $\ZZ$,
  $\WPlZ{\ZZ} P \In \Loss (\YY \shorttimes \ZZ) \Fun \Loss (\XX \shorttimes \ZZ)$,
  is
\expandafter\xlabel[\ref*{eWplSkip}--\ref*{eWplComp}]{eqsWplSkipAbortComp}
\expandafter\xlabel[\ref*{eWplAssign}--\ref*{eWplUnvar}]{eqsWplAssignVarUnvar}
\expandafter\xlabel[\ref*{eWplIf}--\ref*{eWplWhile}]{eqsWplIfWhile}
\expandafter\xlabel[\ref*{eWplIf}--\ref*{eWplPrint}]{eqsWplIfWhilePrint}
\begin{subequations}%
\begin{align}
\WPLZ{\ZZ} \Skip E &= E
\label{eWplSkip}
\\
\WPLZ{\ZZ} {\Abort} E &= \IB{\False}
\label{eWplAbort}
\\
\WPLZ{\ZZ} {(P \mathbin; Q)} E &= \WPLZ{\ZZ} P {(\WPLZ{\ZZ} Q E)}
\label{eWplComp}
\\
\WPLZ{\ZZ} {\progVar{x} \AssignProb f(\progVar{x})} E
  &= \fnMap (f^\intercal \otimes \fnId_\ZZ) E
\label{eWplAssign}
\\
\WPLZ{\ZZ} {\HidVar \progVar{x} \AssignProb f(\progVar{y})} E &=
  \fnMap ((\fnDiag_\YY^\intercal \Comp (\fnId_\YY \otimes f^\intercal)) \otimes \fnId_\ZZ) E
\label{eWplVar}
\\
\WPLZ{\ZZ} {\Unvar \progVar{x}} E &=
  \fnMap (\fnId_\YY \otimes \elemOne_\XX \otimes \fnId_\ZZ) E
\label{eWplUnvar}
\\
\WPLZ{\ZZ} {\If g(\progVar{x}) \Then P \Else Q} E
  &=
    {g^\ZZ \boxtimes \WPLZ{\ZZ} P E} +
    (\lnot g)^\ZZ \boxtimes \WPLZ{\ZZ} Q E
\label{eWplIf}
\\
\label{eWplWhile}
\WPLZ{\ZZ} {\While g(\progVar{x}) \{ P \}} E
  &= \sum_{n \in \Nat}
  \bigl((g^\ZZ \boxtimes -) \Comp \WPlZ{\ZZ} P\bigr)^n \bigl((\lnot g)^\ZZ \boxtimes E\bigr)
\\
\WPLZ{\ZZ} {\Print f(\progVar{x})} E
  &= \sum_{\omega \in \Obsvs} \bigl(f^\intercal \elemOne_{\omega} \otimes \elemOne_\ZZ\bigr) \boxtimes {E}
\label{eWplPrint}
\\
\WPLZ{\ZZ} {P_1 \NonDet P_2} E
  &= \WPLZ{\ZZ}{P_1}{E} \,\MIN\, \WPLZ{\ZZ}{P_2}{E}
\label{eWplNondet}
\end{align}
\end{subequations}
When $\ZZ$ is a singleton set we write $\Wpl$.
\end{definition}

\begin{remark}
  \label{r051342}
  Assignment \Eqn{eWplAssign} applies $\WPe{f}$ pointwise to $E$,
  and \Eqn{eWplVar} and \Eqn{eWplUnvar}
  are analogues that
  ``copy/discard'' parts of the state.
  Nondeterministic choice \Eqn{eWplNondet} is exactly the meet of its
  constituent programs.

  For $\Print$,
  applying \Eqn{e1731a}
  to \Eqn{eWplPrint}
  we find that
  the post-loss $E$
  is evaluated against all possible \emph{choice functions} $h \In \Obsvs \Fun E$
  to give
  the expected minimum \ExpVal\@ of its conjuncts \emph{conditioned} on $\omega$.
  The sums for
  \Eqn{eqsWplIfWhile}
  are similarly ``visible''.
  For $\If$ we ``condition'' (sum) over each branch;
  for $\While$, over each iteration count.
\end{remark}
As with \Ex{ex1713},
let
$\BB$ be a Boolean predicate over $\XX$.
Then $r \mkern1mu \IB{\BB} \sqsubseteq \WPLZ{\ZZ}{P}{E_\YY}$
iff $\forall e_\YY \in E_\YY$
the final \ExpVal\@ of $e_\YY$ is at least $r$
whenever $\BB$ initially holds.

\begin{proposition}[Loops (least fixed point)]
  \label{p1222}
  Fix $P \In \XX \ProgFromTo \XX$.
  Given a loss transformer
  $\alpha \In \Loss (\XX \cramped\shorttimes \ZZ) \Fun \Loss (\XX \cramped\shorttimes \ZZ)$,
  let
  $
    \fnWhileStep(\alpha)(E) =
    g^\ZZ \boxtimes \WPLZ{\ZZ}{P}{\alpha(E)}
    +
    \allowbreak
    (\lnot g)^\ZZ \boxtimes E
  $.
  Then
  \Eqn{eWplWhile} matches the standard
  least fixed point definition
  \Eqn{e051559}:
  \begin{equation}
  \label{e051559}
    \WPlZ{\ZZ} {\While g \{ P \}} =
    \bigl(
      \LeastFixedPoint \alpha \In \bigl(\Loss (\XX \cramped\shorttimes \ZZ) \Fun \Loss (\XX \cramped\shorttimes \ZZ)\bigr)
      \cdot \fnWhileStep(\alpha)
    \bigr)
\end{equation}
\end{proposition}

\begin{example}
  \label{ex061332}
  We have that $\WPLZ{\ZZ}{\Assert g(\progVar{x})}{E} = g^\ZZ \boxtimes E
  = \WPLZ{\ZZ}{(
  \If g(\progVar{x}) \Then \allowbreak \Skip \Else \Abort)}{E}$.
  So assertions
  can be interpreted either as
  hidden assignment or visible control flow.
\end{example}

\begin{example}
  \label{ex061333}
  We reprise \EX{ex052207} using $\Wpl$.
  \settowidth{\dimen128}{$\Wpl.$}
  \begin{align*}
    &
    \mathrel{\hphantom{=}}
    \hspace{-\dimen128}
    \WPL{(\printNDivTwo ; \Var \progVar{b} \AssignProb 0 \NonDet 1)}{\IB{\predNAndBEven}}
    \\
    & =
    \WPL{\mathopen{}\printNDivTwo}{\bigl(
        \IB{\predNEven}
        \MIN
        \IB{\predNOdd}
    \bigr)}
    \\
    & =
    \IB{\progVar{n} \in \{0, 1\}} \boxtimes (\IB{\predNEven} \MIN \IB{\predNOdd})
    +
    \IB{\progVar{n} \in \{2, 3\}} \boxtimes (\IB{\predNEven} \MIN \IB{\predNOdd})
    \\
    & =
    \left(
      \IB{\progVar{n} = 0}
      \MIN
      \IB{\progVar{n} = 1}
    \right)
    +
    \left(
      \IB{\progVar{n} = 2}
      \MIN
      \IB{\progVar{n} = 3}
    \right)
    \\
    & =
    \left( \IB{\progVar{n} \cramped= 0} \cramped+ \IB{\progVar{n} \cramped= 2} \right)
    \MIN
    \left( \IB{\progVar{n} \cramped= 0} \cramped+ \IB{\progVar{n} \cramped= 3} \right)
    \MIN
    \left( \IB{\progVar{n} \cramped= 1} \cramped+ \IB{\progVar{n} \cramped= 2} \right)
    \MIN
    \left( \IB{\progVar{n} \cramped= 1} \cramped+ \IB{\progVar{n} \cramped= 3} \right)
  \end{align*}
\end{example}

\begin{definition}[Refinement, \KuifjeNonDet\@ programs]
\label{d101822}
Given $P, Q \In \XX \ProgFromTo \YY$,
we say $P$ is \underline{\emph{refined by}} $Q$
(``$P \sqsubseteq Q$'')
whenever $\WPLZ{\ZZ}{P}{E} \sqsubseteq \WPLZ{\ZZ}{Q}{E}$ for all $E$.
\end{definition}

\section{Healthiness for \Wpl}
\label{sec205}

For our simulations we will allow abstract programs,
i.e.\@ loss transformers that may not arise as some $\WPl{P}$.
Lemmas \ref{l101552} and \ref{l1610}
give ``healthiness'' (well-formedness) properties
\cite{Dijkstra:76,Back:98,Morgan:96d}
that are obeyed by $\Wpl$,
which we subsequently require for our simulations.

A loss transformer (i.e.\@ function) $f \In \Loss \YY \Fun \Loss \XX$ is
\emph{partial} if $f(\elemOne_\YY) \leq \elemOne_\XX$;
it is \emph{total} if $f(\elemOne_\YY) = \elemOne_\XX$.
It is \emph{homogenous} if $\forall r \In \RExtd$, $r f(E) = f(r E)$.
It is \emph{superlinear} (resp.\@ \emph{linear})
if it is homogenous and $f(E_1 + E_2) \geq f(E_1) + f(E_2)$ (resp.\@ $=$).

\begin{definition}
  \label{d1357}
  A \KuifjeNonDet\@ program
  $P \In \XX \ProgFromTo \YY$
  is \underline{\emph{hidden}} if
  it has no $\If$, $\While$ or $\Print$ instructions.
  It is \underline{\emph{choiceless}} if it has no $\NonDet$ instructions.
\end{definition}

\begin{lemma}
  \label{l101552}
  Let $P \In \XX \ProgFromTo \YY$ be a \KuifjeNonDet\@ program.
  Then
  $\WPlZ{\ZZ}{P} \In \Loss (\YY \shorttimes \ZZ) \Fun \Loss (\XX \shorttimes \ZZ)$ is
  superlinear,
  partial, and
  Scott-cts.
  Additionally,
  if $P$ is \emph{hidden} then
  $\WPlZ{\ZZ}{P}$ is \emph{$\MIN$-preserving};
  if $P$ is \emph{choiceless} then
  $\WPlZ{\ZZ}{P}$ is \emph{linear}.
\end{lemma}

\begin{remark}
  \label{r1526}
  Given a superlinear, Scott-cts.\@ loss transformer $f$,
  finite meets
  (resp.\@ infinite sums)
  such as in \Eqn{eWplNondet}
  (resp.\@ \Eqn{eqsWplIfWhilePrint})
  \emph{oplaxly distribute} (resp. \emph{laxly distribute})
  over $f$,
  so
  $\mathord{\MIN} \Comp f \HRefR f \Comp \mathord{\MIN}$
  and
  $\bigl.\mathord{\sum}\bigr. \Comp f \HRef f \Comp \bigl.\mathord{\sum}\bigr.$ respectively.

  We have distributivity
  exactly when $f$ is
  \MIN-preserving
  (resp.\@ linear),
  hence setting $\rep = f$
  we satisfy
  the left (resp.\@ right) side of
  \Eqn{eSimCosimWeakCommut},
  minding the contravariance of $\Wpl$
  \Eqn{eOPWplSquares}.
  This foreshadows the role played by
  hidden (resp.\@ choiceless) programs
  in our forward (resp.\@ backward) simulation rule.
\end{remark}
Because our datatypes and program contexts may operate on separate parts of the state space,
we require a characterisation of \emph{context extension}.
From \cite[\S5.3]{Chen:25ab}
we derive a \emph{``weak frame rule''}:
\begin{equation}
  \label{e061849}
  \WPLZ{\ZZ}{P}{(e_\ZZ \boxtimes E_\YY)}
  =
  e_\ZZ \boxtimes \WPL{P}{E_\YY}
\end{equation}
Unfortunately, analogously to \Eqn{e1834},
$
  (\Print \progVar{y} ; \progVar{x} \AssignProb 0 \NonDet 1)
  \HRefNeq
  (\progVar{x} \AssignProb 0 \NonDet 1 ; \Print \progVar{y})
$,
so extending $\boxtimes$ to a monoid on $\Loss \XX$
would not give a true frame rule.
Instead, \Lem{l1610} gives ``weak independence''
in terms of
\emph{correlation transformers}
\cite{Rabehaja:19ab}.
\begin{definition}
\label{d101801}
A \underline{\emph{correlation transformer}} $f \In \XX \ProgFromTo \YY$
is a family of loss transformers $f^\ZZ \In \Loss (\YY \shorttimes \ZZ) \rightarrow \Loss (\XX \shorttimes \ZZ)$
(for all state spaces $\ZZ$)
such that
$
    \fnMap (\fnId_\XX \otimes g) \circ f^\ZZ =
    f^\WW \circ \fnMap (\fnId_\YY \otimes g)
  $
for all linear, partial, Scott-cts.\@ $g \In \Pred \ZZ \Fun \Pred \WW$.
\end{definition}

\begin{remark}
  \label{r061724}
  Our correlation transformers are defined more restrictively
  than those of \cite{Rabehaja:19ab},
  which only requires commutativity
  with \emph{total} hidden assignment.
  We claim ours is a \emph{strictly} stronger notion,
  ruling out
  unhealthy abstract programs
  well known in QIF
  \cite[Ex.\@ 4.15]{Alvim:20a}.
\end{remark}

\begin{lemma}
  \label{l1610}
  Given a \KuifjeNonDet\@ program $P \In \XX \ProgFromTo \YY$,
  its various context extensions
  $\WPlZ{\ZZ}{P} \In \Loss(\YY \shorttimes \ZZ) \Fun \Loss(\XX \shorttimes \ZZ)$
  form a correlation transformer.
\end{lemma}

\section{Simulations for $\KuifjeNonDet$}
\label{sec060}

For our simulations (\Def{d071245}),
we must mind that $\Wpl$ is contravariant.
E.g.\@ the center squares of \Eqn{eqsSimCosimCD}
now become
\Eqn{eOPWplSquares}, and so on.

\begin{equation}
\label{eOPWplSquares}
\begin{tikzcd}[
  refinementlabel/.style={scale=1},
  row sep=1.5em,
  column sep=large,
]
  \AA \!\times\! \SS
    \arrow[d,phantom,"" {name="rep0"}]
  &
  \AA \!\times\! \SS
    \arrow[l, "\WPl{\OP_j}" swap]
    \arrow[d,phantom,"" {name="rep1"}]
  \\
  \CC \!\times\! \SS \arrow[u, "\repFwd^\SS"]
  &
  \CC \!\times\! \SS \arrow[u, "\repFwd^\SS" swap]
  \arrow[l, "\WPl{\OP'_j}"]
  \arrow[phantom, from="rep0", to="rep1", "\sqsupseteq" refinementlabel]
\end{tikzcd}
\quad
\begin{tikzcd}[
  refinementlabel/.style={scale=1},
  row sep=1.5em,
  column sep=large,
]
  \AA \!\times\! \SS \arrow[d, "\repBwd^\SS" swap]
    \arrow[d,phantom,"" {name="rep0"}]
  &
  \AA \!\times\! \SS \arrow[d, "\repBwd^\SS"]
    \arrow[l, "\WPlZ{\AA}{\OP_j}" swap]
    \arrow[d,phantom,"" {name="rep1"}]
  \\
  \CC \!\times\! \SS
  &
  \CC \!\times\! \SS
  \arrow[l, "\WPlZ{\CC}{\OP'_j}"]
  \arrow[phantom, from="rep0", to="rep1", "\sqsubseteq" refinementlabel]
\end{tikzcd}
\end{equation}

\subsection{Forward simulations}
\label{sec065}

\begin{figure}[pt]
\begin{gather*}
\setlength\arraycolsep{.5em}
\begin{array}{c|c}
  D_\AA & D_\CC \\ \hline
\begin{aligned}[t]
  I_\AA &\colon \mathopen{}\HidVar \progVar{b} \Assign 0 \PC{\frac{1}{2}} 1  ; \\[0.33em]
  \OP_\AA &\colon \progVar{s} \Assign \progVar{b}; \\[0.33em]
  F_\AA &\colon
    \begin{lgathered}[t]
      \Print \progVar{b}; \\
      \Unvar \progVar{b};
    \end{lgathered}
\end{aligned}
        &
\begin{aligned}[t]
  I_\CC &\colon \mathopen{} \HidVar \progVar{b} \Assign 0 \PC{\frac{1}{2}} 1  ; \\[0.33em]
  \OP_\CC &\colon
    \begin{lgathered}[t]
    \progVar{s} \Assign \progVar{b}; \\
    \Print \progVar{b};
    \end{lgathered} \\[0.33em]
  F_\CC &\colon \mathopen{} \Unvar \progVar{b};
\end{aligned}
\end{array}
\end{gather*}
\caption[(PLACEHOLDER for \textbackslash listoffigures]{%
  Nonrefinement from \Ex{exSimHealth}.
  Shared state is $\HidVar \progVar{s} \In \{0,1\}$.
  \label{figSimUnhealthy}
}
\end{figure}
\begin{example}\label{exSimHealth}
\begin{subequations}
Forward simulations that leak can be unsound.

Consider the datatypes in \Fig{figSimUnhealthy}.
Setting $\CtxP(\OP) = \left(\OP ; (\progVar{s} \Assign \lnot \progVar{s}) \NonDet \Skip\right)$
demonstrates that $D_\AA \not \sqsubseteq D_\CC$.
Yet there is a forward simulation $\rep = \Print \progVar{b}$
satisfying \Eqn{eSimCD}.
In a sense, the $\Print$ in \rep\@
``masks'' the $\Print$ in $\OP_\CC$.
\end{subequations}
\end{example}

\begin{theorem}
  \label{thm1234}
    Let correlation transformer $\rep \In \AA \ProgFromTo \CC$ be a
    forward simulation from $D_\AA$ to $D_\CC$.
    If $\rep$
    is superlinear, partial, Scott-cts.\@,
    and \underline{\MIN-preserving}, then
    $D_\AA \DataRef D_\CC$.
\end{theorem}

\begin{corollary}
  \label{c5050}
  Let $R \In \AA \ProgFromTo \CC$ be a
  \underline{\emph{hidden}}
  \KuifjeNonDet\@ program
  such that $\WPl{R}$ is a
  forward simulation from $D_\AA$ to $D_\CC$.
  Then $D_\AA \sqsubseteq D_\CC$.
\end{corollary}

\subsection{Backward simulations}
\label{sec067}

\begin{figure}[pt]
\begin{gather*}
\setlength\arraycolsep{.5em}
\begin{array}{c|c}
  D_\AA & D_\CC \\ \hline
\begin{aligned}[t]
  I_\AA &\colon \mathopen{} \HidVar \progVar{b} \Assign 0 \NonDet 1; \\[0.33em]
  \OP_\AA &\colon
    \begin{lgathered}[t]
    \progVar{s} \Assign \progVar{b}; \\
    \progVar{b} \Assign 0 \NonDet 1; \\
    \end{lgathered} \\[0.33em]
  F_\AA &\colon \mathopen{} \Unvar \progVar{b};
\end{aligned}
        &
\begin{aligned}[t]
  I_\CC &\colon \mathopen{} \HidVar \progVar{b} \Assign 0 \NonDet 1; \\[0.33em]
  \OP_\CC &\colon
    \begin{lgathered}[t]
    \progVar{b} \Assign 0 \NonDet 1; \\
    \progVar{s} \Assign \progVar{b};
    \end{lgathered} \\[0.33em]
  F_\CC &\colon \mathopen{} \Unvar \progVar{b};
\end{aligned}
\end{array}
\end{gather*}
\caption[(PLACEHOLDER for \textbackslash listoffigures]{%
  Nonrefinement from \Ex{exCosimHealth}.
  Shared state is $\HidVar \progVar{s} \In \{0,1\}$.
  \label{figCosimUnhealthy}
}
\end{figure}
\begin{example}\label{exCosimHealth}
Nondeterministic backward simulations can be unsound.

Consider the datatypes in \Fig{figCosimUnhealthy}.
Setting $\CtxP(\OP) =
\left(\Print \progVar{a} ; \OP ; \progVar{a} \Assign \progVar{s} \Xor \progVar{a}\right)$
demonstrates that $D_\AA \not \DataRef D_\CC$.
Yet there is a backward simulation $\rep = (\progVar{b} \Assign 0 \NonDet 1)$
satisfying \Eqn{eCosimCD}.
In a sense, the $(\NonDet)$ in \rep\@
``masks'' the $(\NonDet)$ in $\OP_\CC$.
\end{example}

\begin{theorem}
  \label{thm1235}
    Let correlation transformer $\rep \In \CC \ProgFromTo \AA$ be a
    backward simulation from $D_\AA$ to $D_\CC$.
    If $\rep$
    is \underline{linear}, partial, and Scott-cts.\@, then
    $D_\AA \DataRef D_\CC$.
\end{theorem}

\begin{corollary}
  \label{c5051}
  Let $R \In \CC \ProgFromTo \AA$ be a
  \underline{\emph{choiceless}}
  \KuifjeNonDet\@ program
  such that $\WPl{R}$ is a
  backward simulation from $D_\AA$ to $D_\CC$.
  Then $D_\AA \sqsubseteq D_\CC$.
\end{corollary}

\section{Examples}
\label{sec150}

\subsection{Example: random bit generator / ``fork+spade''}
\label{sec152}

\begin{example}
\label{ex1538}
Recall the random bit generator example (\Fig{figRandBits}, \Ex{ex091549}).
In \KuifjeNonDet\@
we interpret the encapsulated state as hidden, replacing
``$\Var$'' with ``$\HidVar$'' to get
$I_\CC = ( \HidVar b \Assign 0 \PC{\frac{1}{2}} 1 )$.

Defining a forward simulation $\rep = (\HidVar b \Assign 0 \PC{\frac{1}{2}} 1)$,
we verify \Eqn{eSimCD}:
\begin{align*}
  \Skip ; \rep &\HRef \rep
  \\
  \left(
  \begin{lgathered}
  \progVar{s} \Assign 0 \PC{\frac{1}{2}} 1 ; \\
  \HidVar \progVar{b} \Assign 0 \PC{\frac{1}{2}} 1 ;
  \end{lgathered}
  \right)
  \!
  &\HRef
  \!
  \left(
  \begin{aligned}
    \HidVar \progVar{b} &\Assign 0 \PC{\frac{1}{2}} 1;
    \\
    \progVar{s} &\Assign \progVar{b}; \;
    \progVar{b} \Assign 0 \PC{\frac{1}{2}} 1;
  \end{aligned}
  \right)
  \\
  \Skip &\HRef (\HidVar \progVar{b} \Assign 0 \PC{\frac{1}{2}} 1 ; \Unvar \progVar{b})
\end{align*}
and so \COR{c5050} gives $D_\AA \sqsubseteq D_\CC$ as expected.

In fact, these are all refinement-equalities, so $\rep \In \AA \ProgFromTo \CC$ also serves as a
backward simulation from $D_\CC$ to $D_\AA$.
By \COR{c5051}, $D_\CC \sqsubseteq D_\AA$,
and so the datatypes are equivalent.
\end{example}

\begin{remark}
\label{r1538}
We are not aware of a linear backward simulation
showing $D_\AA \sqsubseteq D_\CC$.
The difficulty is that any $\OP_\CC; \rep$
reveals information about the initial value of $\progVar{b}$
which, though irrelevant, causes \Eqn{eCosimCD} to fail.
\end{remark}

\subsection{Example: encrypted database}
\label{sec155}

\begin{figure}[pt]%
  \def\LoopGuardA{\left(\progVar{n} \neq N \land \progVar{H}[\progVar{n}] \neq \progVar{x}  \right) }
  \def\LoopGuardB{\left(\progVar{n} \neq N \land \progVar{H}[{(\progVar{n}\!+\!\progVar{m})\%N}] \neq \progVar{x}  \right) }
\begin{gather*}
\text{Shared state:} \quad
\HidVar \progVar{x} \In \XX, \; \progVar{U} \In \XX[N], \; \progVar{r} \In \{0,1\}.
\\
\begin{array}{c}
  D_\AA
  \\ \hline
\begin{aligned}[t]
  I_\AA &\colon \HidVar \progVar{H} \AssignProb \progVar{U};
        &
  \OP_\AA &\colon \progVar{r} \Assign (\progVar{x} \in \progVar{H});
        &
  F_\AA &\colon \Unvar \progVar{H};
\end{aligned}
\end{array}
\\[0.33em]
\setlength\arraycolsep{.5em}
\begin{array}{c|c}
  D_\CC & D'_\CC \\ \hline
\begin{aligned}[t]
  I_\CC &\colon \!\HidVar \progVar{H} \AssignProb \progVar{U}; \\[0.33em]
  \OP_\CC &\colon
    \begin{lgathered}[t]
      \HidVar \progVar{n} \Assign 0; \\
      \While \LoopGuardA \{ \\
      \quad \progVar{n} \Assign \progVar{n}+1; \\
      \}; \\
      \progVar{r} \Assign (\progVar{n} < N); \\
      \Unvar \progVar{n};
    \end{lgathered} \\[0.33em]
  F_\CC &\colon \!\Unvar \progVar{H};
\end{aligned}
        &
\begin{aligned}[t]
  I'_\CC &\colon \!\HidVar \progVar{H} \AssignProb \progVar{U}; \\[0.33em]
  \OP'_\CC &\colon
    \begin{lgathered}[t]
      \HidVar \progVar{n} \Assign 0; \\
      \HidVar \progVar{m} \Assign
      0 \PC{\frac{1}{N}} 1 \PC{\frac{1}{N}} \ldots \PC{\frac{1}{N}} N\!-\!1; \\[0.2em]
      \While \LoopGuardB \{ \\
      \quad \progVar{n} \Assign \progVar{n}+1; \\
      \}; \\
      \progVar{r} \Assign (\progVar{n} < N); \\
      \Unvar \progVar{n};
      \Unvar \progVar{m};
    \end{lgathered} \\[0.33em]
  F'_\CC &\colon \!\Unvar \progVar{H};
\end{aligned}
\end{array}
\end{gather*}
\caption[]{%
  An ``encrypted database'' $D_\AA$ \cite[Fig.\@ 6]{Chen:24aab},
  and flawed implementations $D_\CC$ \cite[Fig.\@ 7]{Chen:24aab}
  and $D'_\CC$
  that leak side channel information.

  In $\OP'_\CC$, $\progVar{m}$
  is initialised uniformly over the first $N$ naturals.
  \label{figDatabase}
}
\end{figure}

\Fig{figDatabase}
builds upon an ``encrypted database'' example
from \cite[\S4]{Chen:24aab}.
$D_\AA$ provides membership testing for a secret array,
and $D_\CC$ performs linear search with short-circuiting.
$D_\CC$
is vulnerable to a \emph{timing attack}
leaking the index at which $\progVar{x}$ was found,
hence $D_\AA \not \DataRef D_\CC$ \cite[Eqns.\@ 6--7]{Chen:24aab}.

We might consider the alternative implementation $D'_\CC$ (\Fig{figDatabase}),
which uses a (hidden) random offset to obfuscate
which index is checked at each iteration of the loop.
But in fact this is insufficient to fully hide the database contents.

Let $\XX = \{a,b,c\}$,
$\CtxP(\OP) = (
  \progVar{x} \Assign a; \OP; \allowbreak \Var \progVar{v} \Assign 2 \sqcap 3
)$.
Suppose the array $\progVar{U}$ is initialised uniformly from
$\delta \In \Dist (\XX[4]) = [a,b,c,a] \PC{\frac{1}{2}} [a,b,a,c]$.
As
$\OP_\AA$ leaks nothing about the position of $\progVar{c}$,
we have
$
\WPL{
  \bigl(\progVar{U} \Assign \delta ; I_\AA; \CtxP(\OP_\AA)\bigr)
}{
  \IB{\progVar{\progVar{U}[\progVar{v}] \neq \progVar{c}}}
} \geq \frac{1}{2}
$.
However, this is not the case for $D'_\CC$.
Let
$e_n = \IB{\forall 0 \leq i \leq n \cdot \bigl(
  \progVar{H}[(\progVar{m}+i)\%N] = a
  \Iff
  \allowbreak
  i = n
\bigr)}$.
Then
\begin{align*}
\MoveEqLeft
\WPL{
  (
  \progVar{U} \Assign [a,b,c,a] \PC{\frac{1}{2}} [a,b,a,c] ;
  I'_\CC;
  \CtxP(\OP'_\CC)
  )
}{\IB{\progVar{U}[\progVar{v}] \!\neq\! c}} \\
  &=
\WPL{(
    \progVar{U} \Assign \delta ;
    I'_\CC;
    \progVar{x} \Assign a;
    \OP'_\CC )
  }{
    (
    \IB{\progVar{U}[2] \neq c}
    \MIN
    \IB{\progVar{U}[3] \neq c}
    )
  } \\
  &=
\WPL{\left(
    \begin{multlined}
      \progVar{U} \Assign \delta;
      \Var \progVar{m} \Assign \mathopen{} \\[-3pt]
      {0 \PC{\frac{1}{N}} \ldots \PC{\frac{1}{N}} N\cramped-1}
    \end{multlined}
  \right)}{
    \sum_{n=0}^{N-1}
    \left(
      e_n
      \boxtimes
      (
      \IB{\progVar{U}[2] \neq c}
      \MIN
      \IB{\progVar{U}[3] \neq c}
      )
    \right)
  } \\
  &=
\WPL{\left(
    \begin{multlined}
      \Var \progVar{m} \Assign \mathopen{} \\[-3pt]
      {0 \PC{\frac{1}{N}} \ldots \PC{\frac{1}{N}} N\cramped-1}
    \end{multlined}
\right)}{
  \sum_{n=0}^{N-1}
    e_n
    \boxtimes
    \left(
      \begin{multlined}
        \tfrac{1}{2} \!
        \left(
          {
            \textstyle\sum\nolimits_{U \in \supp \delta}
          }
          \IB{U[2] \neq c}
        \right)
        \MIN
        \\
        \tfrac{1}{2} \!
        \left(
          {
            \textstyle\sum\nolimits_{U \in \supp \delta}
          }
          \IB{U[3] \neq c}
        \right)
      \end{multlined}
    \right)
}
  \\
  &\leq
  \tfrac{1}{4} \bigl(
     ( 1 \sqcap 1 ) +
     ( \tfrac{1}{2} \sqcap 1 ) +
     ( \tfrac{1}{2} \sqcap 0 ) +
     ( 0 \sqcap 0 )
   \bigr)
   = \tfrac{3}{8}
\end{align*}
and so by
  \Eqn{e051246},
$D_\AA \not \sqsubseteq D'_\CC$.

This issue can be resolved with further restrictions on $U$.
Let $e_\mathrm{uniq}$ evaluate to 1 when all elements of $U$ are distinct, and 0 otherwise.
Let $D'_\AA := ((I_\AA; \Assert e_\mathrm{uniq}), \OP_\AA, F_\AA, \AA)$.
In other words, its behaviour is undefined when there are duplicate elements.
Then $\rep := \Assert e_\mathrm{uniq}$
is a backward simulation satisfying the conditions of \Thm{thm1235},
and so
$D'_\AA \sqsubseteq D'_\CC$.

\section{Conclusions and future work}
\label{sec250}

We introduced a ``loss transformer'' semantics for POMDPs.
These POMDPs have a single \NonDet-resolver:
leaks on one side of the encapsulation barrier
affect nondeterminism on the other side.
This is too restrictive
to model e.g.\@
true ``programmer's choice''
or
secure multi-party computation---%
this may require
extending
loss functions
to
multiple resolvers
with \emph{independent knowledge}
\cite{Giro:09ab}
\unskip.

\KuifjeNonDet\@ is not a conservative extension of pGCL (\Rem{r101101}),
in part because all state is hidden.
In future work we intend to
model ``first-class'' visible variables
via \emph{exponentiation} (``$(\Loss \HH)^\VV$''),
and study the semantics of
`promoting' hidden state to visible
(``hypernormalisation'' \cite{Jacobs:17a,Garner:23}).

\smallskip
Our
\emph{healthiness conditions}
for \KuifjeNonDet\@
(%
\Lem{l101552}
\unskip--%
\ref{l1610}
\unskip)
are strong enough to
give our forward (resp.\@ backward) simulation rules
modulo
\MIN-preserving (resp.\@ linearity) conditions.
They
resemble those
for pGCL \cite[\S7]{Morgan:96d}.
However pGCL's are \emph{complete} for
(the image of) \Wpe\@,
whereas
ours
may be
incomplete for $\Wpl$'s
(even after ``dcpo-closure''/sobrification).

\smallskip
Our forward and backward simulations
for \KuifjeNonDet\@ data refinement
are \emph{sound}
modulo ``hidden/choiceless'' restrictions
(\Thm{thm1234}--\ref{thm1235}).
We leave \emph{completeness} to future work.
Forward simulation
is incomplete, even for GCL \cite{Gardiner:93}.
We suspect
backward simulation is also
incomplete for \KuifjeNonDet\@
(\Rem{r1538}).

Backward simulation composed with forward is \emph{jointly} complete for GCL
\cite{He:86ab,Hoare:87,Lynch:95}.
But it is unclear how
to perform
standard constructions
such as
limits of Kan extensions
\cite{He:86ab,Gardiner:93}
or
final coalgebras
\cite{Lynch:95,Hasuo:06}
while preserving
superlinearity, Scott-cty.,
or
(as required)
linearity
or
\MIN-preservation.

Additionally, the requirement that a family of $\rep^\ZZ$ be a correlation transformer
(\Def{d101801})
is difficult to establish in the absence of a corresponding syntactic \KuifjeNonDet\@ program.
It is an open question under what circumstances
a function $\Loss \YY \Fun \Loss \XX$
``extends'' to a correlation transformer.

\begin{credits}

\subsubsection{\discintname}
The authors have no competing interests to declare that are
relevant to the content of this article.
\end{credits}

\printbibliography

\ifAPPENDIX
\appendix

\section{Deferred proofs}

Many of the following proofs use structural induction
over
$\KuifjeNonDet$ programs or program contexts.
We justify well-foundedness in \S\ref{sec345}.

\begin{proof}[\Prop{p1222}]
  Given a dcpo $S$,
  write
  $\Endo[S]$
  for the dcpo
  Scott-cts.\@ endofunctions
  $\Endo[S]
  :=
  \setcmp{\alpha \In S \Fun S
  \!}{\!
    \alpha \; \text{Scott-cts.}
  }$.
  It is well-known
  that
  $\Endo[S]$
  is also a dcpo
  with pointwise directed joins
  \cite[Prop.\@ 2.1.18]{Abramsky:94ab}.

  Let $g \leq \elemOne_\XX, P \In \XX \ProgFromTo \XX$
  (per \Fig{figSyntax})
  and define
  $\fnWhileStep$
  as in \Prop{p1222}.
  By
  \Prop{p101646}
  and
  \Lem{l101552},
  $\fnWhileStep(\alpha)$ is Scott-continuous
  whenever $\alpha$ is.
  \unskip\footnote{Note that \Lem{l101552} can be proven without reliance on \Prop{p1222}.}
  Hence
  $\fnWhileStep$
  restricts to an endofunction
  $\fnWhileStep \In \Endo[\Loss(\XX \shorttimes \ZZ)] \Fun \Endo[\Loss(\XX \shorttimes \ZZ)]$.
  Furthermore
  by pointwise $\dirjoin$
  this restriction
  is itself Scott-cts:
  $\fnWhileStep \in \Endo\bigl[\Endo[\Loss(\XX \shorttimes \ZZ)]\bigr]$.

  Recall the following motivating property of dcpos \cite{Markowsky:76}:
  given
  a dcpo $S$ with least element,
  the least fixed point of any
  $f \In \Endo[S]$
  is given constructively by
  $(\LeastFixedPoint s \In S \cdot f(s)) = \dirjoin[n \in \Nat] \unskip \, f^n(\bot)$
  \cite[Thm.\@ 2.1.19]{Abramsky:94ab}.
  For $\fnWhileStep$
  the least element
  is simply
  $\bigwedge \{f \In \Endo[\Loss(\XX \shorttimes \ZZ)]\}
  =
  \fnConstZero = \WPlZ{\smash\ZZ}{\Abort}$.
  Hence we may solve
  for
  $(\LeastFixedPoint \alpha \cdot \fnWhileStep)$
  \unskip---%
  the RHS of \Eqn{e051559}
  \unskip---%
  directly:
  \begin{align*}
    \fnWhileStep^0 . \fnConstZero . E
    &=
      \elemZero
    \\
    \fnWhileStep^{n+1} . \fnConstZero . E
    &=
    \begin{lgathered}[t]
      g^\ZZ \boxtimes \WPLZ{\ZZ}{P}{( \fnWhileStep^{n} . \fnConstZero . E )}
      \\
      \mathclose{}
      +
      (\lnot g)^\ZZ \boxtimes E
    \end{lgathered}
    \\
  \shortintertext{and by induction,}
    \fnWhileStep^n . \fnConstZero . E
    &=
    \sum_{\mathclap{0 \leq m < n}}
    \bigl((g^\ZZ \boxtimes -) \Comp \WPlZ{\ZZ} P\bigr)^m \bigl((\lnot g)^\ZZ \boxtimes E\bigr)
  \end{align*}
  Taking
  $
    \bigl(
      \LeastFixedPoint \alpha \In \Endo[\Loss (\XX \cramped\shorttimes \ZZ)]
      \cdot \fnWhileStep(\alpha)
    \bigr)
    =
     \dirjoin[n \in \Nat] \unskip \, \fnWhileStep^n . \fnConstZero
  $
  gives the
  infinite sum
  definition
  \Eqn{eWplWhile} as desired.
\end{proof}

\subsection{...for healthiness (\S\ref{sec205})}

\begin{proof}[\Lem{l101552}]
  We proceed by structural induction.

  $\WPlZ{\ZZ}{\Skip}$
  \Eqn{eWplSkip}
  is the identity function
  on $\Loss (\XX \shorttimes \ZZ)$,
  and
  $\WPlZ{\ZZ}{\Abort}$
  \Eqn{eWplAbort}
  is the constant $\IB{\False}$ function.
  Recalling that $\IB{\False} = 0 \cdot E$
  is
  the additive identity
  and
  bottom element,
  we see that both
  functions are
  (super)linear, partial, Scott-cts.\@, and
  \MIN-preserving.

  Sequential composition \Eqn{eWplComp}
  is function composition.
  (Super)linearity,
  partiality,
  Scott-continuity,
  and $\MIN$-preservation
  are straightforwardly closed under composition.

  Next we turn to assignment.
  The tensor of linear Scott-cts.\@ functions
  is linear and Scott-cts.\@
  \Eqn{e101604};
  the tensor preserves partiality (\Prop{p1628}).
  Then by \Prop{p1250} and \Prop{p101646},
  each $\fnMap$
  in \EQns{eqsWplAssignVarUnvar}
  is linear, partial, Scott-cts,
  and \MIN-preserving.

  \smallskip
  Next are the
  ``visible''
  (non-hidden)
  constructs
  \Eqn{eqsWplIfWhilePrint}.
  Here we need not prove $\MIN$-preservation.

  We first recall that
  addition and $\RExtd$-multiplication are
  (jointly) Scott-cts.\@,
  and so
  (1) addition (of loss transformers) preserves Scott-continuity,
  and
  (2) directed joins preserve (super)linearity.
  By inductive hypothesis,
  the $\Wpl$'s for
  $\If$, $\While$, and $\Print$
  \Eqn{eqsWplIfWhilePrint}
  are superlinear and Scott-cts.
  If the overall program is choiceless,
  $\If$ and $\While$
  give linear \Wpl's,
  and so does
  $\Print$ (which is a choiceless base case).

  As for partiality,
  by \Prop{p1750}
  it suffices to show that
  the sum of the guards
  $\sum_i g_i \leq \elemOne_\XX$
  (rearranging
  \Eqn{eqsWplIfWhilePrint}
  in the form
  $\sum_i \WPlZ{\ZZ}{\Assert g_i ; P_i}$).
  For $\If$, we have $g + \lnot g = \elemOne_\XX$
  by definition.
  For $\While$,
  we note that the $n$th summand
  is $\leq g^n \boxtimes \lnot g$
  (with $g^n$ indicating powers
  in the conjunction monoid
  $(\Pred \XX, \boxtimes, \elemOne_\XX)$).
  For $\Print$,
  we use the partiality of $f^\intercal$.

  \smallskip
  Finally we consider nondeterminism.
  Here we need not prove linearity.
  Recall that $\MIN$
  is the meet for $\Loss \XX$;
  this immediately means it
  preserves superlinearity, partiality,
  and \MIN-preservation.
  It is well-known
  that $\MIN$ is Scott-cts.\@
  \cite[Lem.\@ 4.21]{Tix:09}.
  Then by inductive hypothesis,
  the $\Wpl$ in \Eqn{eWplNondet}
  is superlinear, partial, and Scott-cts.\@;
  furthermore,
  if the overall program is hidden then
  the $\Wpl$ is meet-preserving.
\end{proof}

\medskip

\begin{proof}[\Lem{l1610}]
  We proceed by structural induction.
  The cases for
  $\Skip$ \Eqn{eWplSkip}
  and
  sequential composition \Eqn{eWplComp}
  are trivial.
  For $\Abort$ \Eqn{eWplAbort} we remark that
  $\IB{\False}$ is absorbing for $(\boxtimes)$.
  Hidden assignment \Eqn{eqsWplAssignVarUnvar}
  follows from
  the functoriality of the tensor
  \Eqn{e1622b}.
  Recalling \Prop{p101646},
  for visible programs
  \Eqn{eqsWplIfWhilePrint}
  we use the linearity and Scott-cty.\@ of
  $\fnMap(\fnId_\XX \otimes g)$;
  for nondeterminism
  \Eqn{eWplNondet},
  we use the \MIN-preserving property of
  $\fnMap(\fnId_\XX \otimes g)$.
\end{proof}

\subsection{...for simulations (\S\ref{sec060})}

In the following, fix
datatypes
$D_\AA = (I, \OP, F, \AA)$ and
$D_\CC = (I', \OP', F', \CC)$
with
signature $(\SS, J)$.

Note that after ``inlining''
a datatype
with
a
program context $\CtxP \In \XX \ProgFromTo \YY$,
the types
become
$\CtxP(\OP) \In
\XX \shorttimes \AA
\ProgFromTo
\YY \shorttimes \AA
$ and
$\CtxP(\OP') \In
\XX \shorttimes \CC
\ProgFromTo
\YY \shorttimes \CC
$
\unskip.

\begin{proof}[\Thm{thm1234}]
  We follow \cite{Gardiner:93} and show that,
  for all program contexts $\CtxP \In \XX \ProgFromTo \YY$ with signature $(\SS, J)$,
  \begin{equation}
    \WPlZ{\ZZ}{\CtxP(\OP)} \Comp \rep^{\YY \shorttimes \ZZ}
    \HRef
    \rep^{\XX \shorttimes \ZZ} \Comp \WPlZ{\ZZ}{\CtxP(\OP')}
    \label{e1652}
  \end{equation}
  We proceed by structural induction.

  \paragraph{Base case ($\OP_j$).}
  Follows from the middle square of \Eqn{eSimCD}.

  \paragraph{Base case ($\Skip$, $\Abort$, hidden assignment).}
  Follows from \Def{d101801}
  and \Lem{l101552}.

\paragraph{Inductive case (composition).}
  \begin{align*}
  \MoveEqLeft
    \WPlZ{\ZZ}{\bigl(\CtxP(\OP); \CtxQ(\OP)\bigr)}
    \Comp
    \rep^{\WW \shorttimes \ZZ}
  \\
  \tag*{[\Eqn{eWplComp}]}
  &=
  \WPlZ{\ZZ}{\CtxP(\OP)}
    \Comp
    \WPlZ{\ZZ}{\CtxQ(\OP)}
    \Comp
    \rep^{\WW \shorttimes \ZZ}
  \\
  \tag*{[ind.\@ hypothesis]}
  &\HRef
  \WPlZ{\ZZ}{\CtxP(\OP)}
    \Comp
    \rep^{\YY \shorttimes \ZZ}
    \Comp
    \WPlZ{\ZZ}{\CtxQ(\OP')}
  \\
  \tag*{[ind.\@ hypothesis]}
  &\HRef
  \rep^{\XX \shorttimes \ZZ}
    \Comp
    \WPlZ{\ZZ}{\CtxP(\OP')}
    \Comp
    \WPlZ{\ZZ}{\CtxQ(\OP')}
  \\
  &=
    \rep^{\XX \shorttimes \ZZ}
    \Comp
    \WPlZ{\ZZ}{\bigl(\CtxP(\OP'); \CtxQ(\OP')\bigr)}
  \end{align*}

\paragraph{Inductive case ($\If$ and $\Print$).}
  Recall (\Ex{ex061332})
  that
  $\WPLZ{\ZZ}{\Assert g}{E} = g^\ZZ \boxtimes E$.
  Observe that
    \Eqn{e1652} inducts over infinite sums of guarded programs:
  \begin{align*}
  \MoveEqLeft
    \left(\sum_{i \in I}
    \WPlZ{\ZZ}{\bigl(\Assert g_i ; \CtxP_i(\OP) \bigr)}\right) \Comp \rep^{\YY \shorttimes \ZZ}
  \\
  &=
  \tag*{[ptwise.\@ addition and $\dirjoin$]}
    \sum_{i \in I}
    \WPlZ{\ZZ}{\bigl(\Assert g_i ; \CtxP_i(\OP) \bigr)} \Comp \rep^{\YY \shorttimes \ZZ}
  \\
  \tag*{[ind.\@ hyp.\@]}
  &\sqsubseteq
    \sum_{i \in I}
    \WPlZ{\ZZ}{\Assert g_i} \Comp \rep^{\XX \shorttimes \ZZ} \Comp \WPlZ{\ZZ}{\CtxP_i(\OP')}
  \\
  \tag*{[corr.\@ txfmr.\@]}
  &=
    \sum_{i \in I}
    \rep^{\XX \shorttimes \ZZ} \Comp
    \WPlZ{\ZZ}{\bigl(\Assert g_i ; \CtxP_i(\OP')\bigr)}
  \\
  &=
  \tag*{[inf.\@ sum \Eqn{eConeInfiniteSum}]}
    \dirjoin[\substack{J \subseteq I \\ \mathclap{J\,\text{finite}}}]
    \sum_{i \in J}
    \rep^{\XX \shorttimes \ZZ} \Comp
    \WPlZ{\ZZ}{\bigl(\Assert g_i ; \CtxP_i(\OP')\bigr)}
  \\
  \tag*{[superlinear]}
  &\sqsubseteq
    \dirjoin[\substack{J \subseteq I \\ \mathclap{J\,\text{finite}}}]
    \rep^{\XX \shorttimes \ZZ} \Comp
    \sum_{i \in J}
    \WPlZ{\ZZ}{\bigl(\Assert g_i ; \CtxP_i(\OP')\bigr)}
  \\
  \tag*{[Scott-cts.\@]}
  &=
    \rep^{\XX \shorttimes \ZZ} \Comp
    \dirjoin[\substack{J \subseteq I \\ \mathclap{J\,\text{finite}}}]
    \sum_{i \in J}
    \WPlZ{\ZZ}{\bigl(\Assert g_i ; \CtxP_i(\OP')\bigr)}
  \\
  \tag*{[inf.\@ sum \Eqn{eConeInfiniteSum}]}
  &=
    \mathop{}%
    \rep^{\XX \shorttimes \ZZ} \Comp
    \sum_{i \in I}
    \WPlZ{\ZZ}{\bigl(\Assert g_i ; \CtxP_i(\OP')\bigr)}
  \end{align*}
  For conditionals
  $\If g(\progVar{x}) \Then \CtxP \Else \CtxQ$, set $I=\{\True,\False\}$ and $g_\True=g, g_\False=\lnot g, P_\True=P, P_\False=Q$.
  For leaks $\Print f(\progVar{x})$ with $f \In \XX \Fun \DistPartial I$, set $g_i=f^\intercal(\elemOne_i), P_i = \Skip$.

\paragraph{Inductive case ($\While$).}
We use the $\LeastFixedPoint$ characterisation
\Eqn{e051559}.
Given program context
$\While g(\progVar{x}) \{ \CtxP \}$,
define a transformer $\fnWhileStep'$ acting on program contexts:
\begin{gather}
  \label{eWhileStepPrime}
  \fnWhileStep'(\CtxQ) =
  \bigl(\If g(\progVar{x}) \{ \CtxP ; \CtxQ \}
  \Else \Skip\bigr)
\intertext{
  Clearly
}
  \label{e1543}
  \WPlZ{\ZZ}{\bigl(\fnWhileStep'(\CtxQ)(-)\bigr)}
  =
  \fnWhileStep\bigl(\WPlZ{\ZZ}{\CtxQ}(-)\bigr)
  \intertext{
and by the same explicit construction as
with \Prop{p1222},
}
  \label{e1329}
  \WPlZ{\ZZ}{\While g(\progVar{x}) \{ \CtxP(-) \}}
  =
  \dirjoin[n \In \Nat]
  \WPlZ{\ZZ}{\left(\fnWhileStep'^n(\Abort)\right)(-)}
\end{gather}
Then
\begin{align*}
  \MoveEqLeft
  \bigl(
    \WPlZ{\ZZ}{\While g(\progVar{x}) \{ \CtxP(\OP) \}}
  \bigr)
  \Comp \rep^{\YY \shorttimes \ZZ}
  \\
  &=
  \dirjoin[n \In \Nat]
  \bigl(
    \WPlZ{\ZZ}{\left(\fnWhileStep'^n(\Abort)(\OP)\right)}
    \Comp \rep^{\YY \shorttimes \ZZ}
  \bigr)
  \\
  \tag*{[ind.\@ hypothesis]}
  &\HRef
  \dirjoin[n \In \Nat]
  \bigl(
    \rep^{\XX \shorttimes \ZZ} \Comp
    \WPlZ{\ZZ}{\left(\fnWhileStep'^n(\Abort)(\OP')\right)}
  \bigr)
  \\
  \tag*{[Scott-cts.]}
  &=
  \rep^{\XX \shorttimes \ZZ} \Comp
  \bigl(
    \WPlZ{\ZZ}{\While g(\progVar{x}) \{ \CtxP(\OP') \}}
  \bigr)
\end{align*}
The inequality follows
pointwise
from the inductive hypothesis.

\paragraph{Inductive case (nondeterminism).}
  \begin{align*}
    \MoveEqLeft
    \WPlZ{\ZZ}{\bigl(\CtxP_1(\OP) \MIN \CtxP_2(\OP)\bigr)} \circ \rep^{\YY \shorttimes \ZZ}
    \\
    \tag*{[\Eqn{eWplNondet}]}
    &=
    \bigl(\WPlZ{\ZZ}{\CtxP_1(\OP)} \MIN \WPlZ{\ZZ}{\CtxP_2(\OP)}\bigr) \circ \rep^{\YY \shorttimes \ZZ}
    \\
    \tag*{[ptwise.\@ $\MIN$]}
    &\HRef
    \bigl(\WPlZ{\ZZ}{\CtxP_1(\OP)} \Comp \rep^{\YY \shorttimes \ZZ}\bigr)
    \MIN
    \bigl(\WPlZ{\ZZ}{\CtxP_2(\OP)} \Comp \rep^{\YY \shorttimes \ZZ}\bigr)
    \\
    \tag*{[ind.\@ hypothesis]}
    &\HRef
    \bigl(\rep^{\XX \shorttimes \ZZ} \Comp \WPlZ{\ZZ}{\CtxP_1(\OP')}\bigr)
    \MIN
    \bigl(\rep^{\XX \shorttimes \ZZ} \Comp \WPlZ{\ZZ}{\CtxP_2(\OP')}\bigr)
    \\
    \tag*{[$\MIN$-preserving]}
    &=
    \rep^{\XX \shorttimes \ZZ} \circ \bigl(
      \WPlZ{\ZZ}{\CtxP_1(\OP')}
      \MIN
      \WPlZ{\ZZ}{\CtxP_2(\OP')}
    \bigr)
  \end{align*}

  \smallskip
  By induction,
  \EQn{e1652}
  holds for all $\CtxP$.
  This
  together with
  the left and right squares of \Eqn{eSimCD}
  gives
  $I; \CtxP(\OP); F \HRef I'; \CtxP(\OP'); F'$ as desired.
\qed
\end{proof}

\bigskip
\bigskip

\begin{proof}[\Thm{thm1235}]
  We show that for all program contexts $\CtxP$ with signature $(\SS, J)$,
  \begin{equation}
    \rep^{\XX \shorttimes \ZZ} \Comp \WPlZ{\ZZ}{\CtxP(\OP)}
    \HRef
    \WPlZ{\ZZ}{\CtxP(\OP')} \Comp \rep^{\YY \shorttimes \ZZ}
    \label{e1455}
  \end{equation}
  We proceed by structural induction.

  \paragraph{Base case ($\OP_j$).}
  Follows from the middle square of \Eqn{eCosimCD}.

  \paragraph{Base case ($\Skip$, $\Abort$, hidden assignment).}
  Follows from \Def{d101801}
  and \Lem{l101552}.

\paragraph{Inductive case (composition).}
\begin{align*}
  \MoveEqLeft
  \rep^{\XX \shorttimes \ZZ} \Comp \WPlZ{\ZZ}{\bigl(\CtxP(\OP) ; \CtxQ(\OP)\bigr)}
  \\
  \tag*{[ind.\@ hypothesis]}
  &\HRef
  \WPlZ{\ZZ}{\CtxP(\OP')}
    \Comp
    \rep^{\YY \shorttimes \ZZ}
    \Comp
    \WPlZ{\ZZ}{\CtxQ(\OP)}
  \\
  &\HRef
  \WPlZ{\ZZ}{\bigl(\CtxP(\OP') ; \CtxQ(\OP')\bigr)}
  \Comp \rep^{\WW \shorttimes \ZZ}
\end{align*}

\paragraph{Inductive case ($\If$ and $\Print$).}
As with \Thm{thm1234}, we argue:
\begin{align*}
  \MoveEqLeft
  \rep^{\XX \shorttimes \ZZ}
  \Comp \sum_{i \in I}
  \WPlZ{\ZZ} {(\Assert g_i ; \CtxP_i(\OP))}
  \\
  \tag*{[linear, Scott-cts.]}
  &=
  \sum_{i \in I}
  \rep^{\XX \shorttimes \ZZ}
  \Comp
  \WPlZ{\ZZ} {(\Assert g_i ; \CtxP_i(\OP))}
  \\
  \tag*{[corr.\@ txfmr.\@]}
  &=
  \sum_{i \in I}
  \WPlZ{\ZZ} {(\Assert g_i)}
  \Comp
  \rep^{\XX \shorttimes \ZZ}
  \Comp
  \WPlZ{\ZZ} {\CtxP_i(\OP)}
  \\
  \tag*{[ind.\@ hypothesis]}
  &\sqsubseteq
  \sum_{i \in I}
  \WPlZ{\ZZ} {\bigl(\Assert g_i ; \CtxP_i(\OP')\bigr)}
  \Comp
  \rep^{\YY \shorttimes \ZZ}
\end{align*}

\paragraph{Inductive case ($\While$).}
As with \Thm{thm1234}, we construct a program transformer $\fnWhileStep'$
per
\Eqn{eWhileStepPrime}.
Then:
\begin{align*}
  \MoveEqLeft
  \rep^{\XX \shorttimes \ZZ}
  \Comp
  \bigl(
    \WPlZ{\ZZ}{\While g(\progVar{x}) \{ \CtxP(\OP) \}}
  \bigr)
  \\
  &=
  \tag*{[$\rep$ Scott-cts.]}
  \dirjoin[n \In \Nat]
  \left(
    \rep^{\XX \shorttimes \ZZ}
    \Comp
    \WPlZ{\ZZ}{\left(\fnWhileStep'^n(\Abort)(\OP)\right)}
  \right)
  \\
  &\HRef
  \tag*{[ind.\@ hypothesis]}
  \dirjoin[n \In \Nat]
  \left(
    \WPlZ{\ZZ}{\left(\fnWhileStep'^n(\Abort)(\OP')\right)}
    \Comp
    \rep^{\YY \shorttimes \ZZ}
  \right)
  \\
  &=
  \tag*{[ptwise.\@ $\dirjoin$]}
  \dirjoin[n \In \Nat]
  \left(
    \WPlZ{\ZZ}{\left(\fnWhileStep'^n(\Abort)(\OP')\right)}
  \right)
  \Comp
  \rep^{\YY \shorttimes \ZZ}
  \\
  &=
  \bigl(
    \WPlZ{\ZZ}{\While g(\progVar{x}) \{ \CtxP(\OP') \}}
  \bigr)
  \Comp
  \rep^{\YY \shorttimes \ZZ}
\end{align*}

\paragraph{Inductive case (nondeterminism).}
  \begin{align*}
    \MoveEqLeft
    \rep^{\XX \shorttimes \ZZ} \Comp \bigl(\WPlZ{\ZZ}{\left(\CtxP_1(\OP) \MIN \CtxP_2(\OP)\right)} \bigr)
    \\
    \tag*{[\Eqn{eWplNondet}]}
    &=
    \rep^{\XX \shorttimes \ZZ} \Comp \bigl(\WPlZ{\ZZ}{\CtxP_1(\OP)} \MIN \WPlZ{\ZZ}{\CtxP_2(\OP)} \bigr)
    \\
    \tag*{[monotone]}
    &\HRef
    \bigl( \rep^{\XX \shorttimes \ZZ} \Comp \WPlZ{\ZZ}{\CtxP_1(\OP)} \bigr)
    \MIN
    \bigl( \rep^{\XX \shorttimes \ZZ} \Comp \WPlZ{\ZZ}{\CtxP_2(\OP)} \bigr)
    \\
    \tag*{[ind.\@ hypothesis]}
    &\HRef
    \bigl( \WPlZ{\ZZ}{\CtxP_1(\OP')} \Comp \rep^{\YY \shorttimes \ZZ} \bigr)
    \MIN
    \bigl( \WPlZ{\ZZ}{\CtxP_2(\OP')} \Comp \rep^{\YY \shorttimes \ZZ} \bigr)
    \\
    \tag*{[ptwise.\@ $\MIN$]}
    &\HRef
    \WPlZ{\ZZ}{\bigl( \CtxP_1(\OP') \MIN \CtxP_2(\OP') \bigr)} \Comp \rep^{\YY \shorttimes \ZZ}
  \end{align*}

  \smallskip
  By induction,
  \EQn{e1455}
  holds for all $\CtxP$.
  This
  together with
  the left and right squares of \Eqn{eCosimCD}
  gives
  $I; \CtxP(\OP); F \HRef I'; \CtxP(\OP'); F'$ as desired.
\qed
\end{proof}

\section{Well-foundedness of induction}
\label{sec345}

Given
a program context $\CtxP$ with signature $(\SS, J)$
(or w.l.o.g.\@ a $\KuifjeNonDet$ program $P$),
assign it an ordinal ``$\ordinalFor{\CtxP}$'' as follows:
\begin{subequations}%
\begin{gather}
  \ordinalFor{\OP_j}
  =
  \ordinalFor{\Skip}
  =
  \ordinalFor{\Abort}
  =
  \ordinalFor{\Print f(\progVar{x})}
  =
  0
  \\
  \ordinalFor{\progVar{x} \AssignProb f(\progVar{x})}
  =
  \ordinalFor{\HidVar \progVar{x} \AssignProb f(\progVar{y})}
  =
  \ordinalFor{\Unvar \progVar{x}}
  =
  0
  \\
  \label{eOrdinalForCompNonDetIf}
  \ordinalFor{\CtxP ; \CtxQ}
  =
  \ordinalFor{\CtxP \NonDet \CtxQ}
  =
  \ordinalFor{\If g(\progVar{x}) \Then \CtxP \Else \CtxQ}
  =
  \ordinalFor{\CtxP} + \ordinalFor{\CtxQ} + 1
  \\
  \ordinalFor{\While g(\progVar{x}) \, \{ \CtxP \}}
  =
  \ordinalFor{\CtxP} \times \omega + \omega
\end{gather}
\end{subequations}
\xlabel[\theequation]{eqsOrdinalFor}%
where
$
\alpha + \beta =
\max\bigl(\alpha, \dirjoin[\gamma < \beta] \! \fnOrdinalSucc(\alpha + \gamma)\bigr)
$
and
$
\alpha \times \beta =
\dirjoin[\gamma < \beta] \! (\alpha \times \gamma + \alpha)
$.

For iteration,
we have the following facts
about
the construction
\Eqn{eWhileStepPrime}:
\begin{align}
  \ordinalFor{\fnWhileStep'(\CtxQ)}
  &=
  \ordinalFor{\CtxP}
  +
  \ordinalFor{\CtxQ}
  +
  2
  \\
  \ordinalFor{\fnWhileStep'^n(\CtxQ)}
  &=
  \ordinalFor{\CtxP} \times n
  +
  \ordinalFor{\CtxQ}
  +
  2 \times n
  \\
  \ordinalFor{\fnWhileStep'^n(\Abort)}
  &=
  \ordinalFor{\CtxP} \times n
  +
  2 \times n
  <
  \ordinalFor{\While g(\progVar{x}) \, \{ \CtxP \}}
  \\
  \ordinalFor{\CtxP}
  &<
  \ordinalFor{\While g(\progVar{x}) \, \{ \CtxP \}}
\end{align}
By the last two inequalities,
$\ordinalFor{\CtxP}$
gives a sensible well-ordering for the prior proofs.

\section{Predicates and loss functions}
\label{sec100}

Our language semantics (\S\ref{sec200}) will be given in terms
of ``weakest precondition'' / ``predicate transformers''.
As a precursor, we consider the spaces of ``predicates'':
the probabilistic predicates $\Pred \XX$ and
the loss functions $\Loss \XX$ (\Def{d071224}).
We will see these are both instances of
the \emph{$d$-cones}
first studied by \cite{Tix:09ab}.

Recall that under the \emph{Scott topology} for a poset,
the closed sets are exactly those closed under directed joins (wherever they exist)
  \cite[Def.\@ 2.3.1]{Abramsky:94ab}.
Hence a \emph{Scott-continuous} function preserves directed joins.

\begin{example}
  Scott-cts.\@ functions are necessarily monotone
  and preserve limits of ascending chains.
  They need not preserve (nonempty) meets,
  e.g.\@ let $f \In \RExtd \Fun \RExtd$ with
  $f(x) = \IB{x \neq 0}$.
\end{example}

\subsection{$d$-cones}
\label{sec840}

We briefly review
the theory of
ordered cones
and
d-cones.
More detailed exposition can be found in
\cite{Tix:09ab,Keimel:17}.

\begin{definition}[{\cite[\S2.1]{Tix:09ab}}]
\label{d101414}%
An \underline{\emph{ordered cone}}
$(C, \leq, +, \elemZero, \cdot)$
is a semiring module for $(\RExtd, 0, +, 1, \times)$,%
\footnote{%
I.e.\@ a set $C$ equipped with a commutative ``addition'' monoid $(C, +, \elemZero)$
and a ``scalar multiplication'' $(\cdot) \In~ \RExtd \times C \Fun C$
that is a monoid action for $(\RExtd, \times, 1)$,
so that $(.)$ distributes over $(+)$.%
}%
\@ equipped with a partial order $(C, \leq)$ such that
addition and scalar multiplication are monotone in both variables.
(It follows that $\elemZero$ is the least element.)

An ordered cone is a \underline{\emph{d-cone}} if it has all directed joins
(and hence a top element $\infty$),
and all operations are Scott-continuous.

A \underline{\emph{continuous d-cone}} is a d-cone that is a continuous domain in the usual
domain theoretic sense. (I.e.\@ all elements are the directed join of their way-belows.)

A morphism of ordered cones
is a linear, monotone function;
a morphism of d-cones
is a linear, Scott-continuous function.
\end{definition}

\begin{example}
  $\RExtd$ is itself a continuous d-cone.
\end{example}
Because the additive identity $\elemZero$
is the least element in a
d-cone,
addition is expansive ($c \leq c + c'$)
and so
infinite sums are well-defined:
\begin{equation}\label{eConeInfiniteSum}
  \sum_{i \in I} c_i =
  \!\!
  \mathop{\hphantom{^\uparrow}\dirjoin}\limits_{\smash{\substack{J \subseteq I \\ J\,\text{finite}}}}
    \sum_{j \in J} c_j
\end{equation}

\subsection{Probabilistic predicates}
\label{sec850}

Given a
(discrete) state space $\XX$
(i.e.\@ a set),
recall (\S\ref{sec070})
that the \emph{probabilistic predicates} over $\XX$
are
$\Pred \XX = (\XX \Fun \RExtd)$.
$\Pred \XX$ is a continuous d-cone
\cite[Prop.\@ 2.25]{Tix:09ab};
the
bounded linear combinations
of ``simple functions''
($\sum_{i=1}^n r_i \elemOne_{x_i\!}$
where
$n \in \Nat$
and
$r_i \In \RealNonNeg, x_i \In \XX$ for $i = 1,\ldots,n$).
\unskip\footnote{
Up to d-cone isomorphism,
the (continuous) d-cone $\Pred \XX$ is a special case
of the \emph{lower semicontinuous functionals} over
a (core compact) topological space,
$
\Pred (S, \OO_S) =
\setcmp{f \In S \Fun \RExtd}{f \, \text{Scott-cts.\@}}
$
\cite{Jones:89ab,Tix:09ab}.
}

Given $e \In \Pred \YY$,
and a linear Scott-cts.\@ function
$f \In \Pred \YY \Fun \Pred \XX$,
\begin{align}
  \label{e061314}
  e = \sum_{y \in \YY} e(y) \shorttimes \elemOne_y
  \; &\overset{\mathclap{\Eqn{eConeInfiniteSum}}}{=} \;
  \dirjoin
    [\mathclap{\substack{F \subseteq \YY, \\ F\,\text{finite}}}] \;
  \sum_{y \in F} e(y) \shorttimes \elemOne_y
  \\
  \label{e061315}
  f(e)(x) &=
  \sum_{y \in \YY}
  e(y) \times f(\elemOne_y)(x)
\end{align}
By \Eqn{e061315}, $f$ is uniquely determined by
$f(\elemOne_y)$ for all $y \In \YY$.
Indeed,
adapting
the argument of
\cite[Prop.\@ 13]{Jacobs:17}
(quoting
\cite[Thm.\@ 2.4.5]{Furber:17}
in turn)
to d-cones:
\begin{proposition}
  \label{p061351}
  There is a
  d-cone isomorphism
  \Eqn{e061445}
  given by
  mapping
  $f(e)(x) = {\textstyle\sum_{y \in \YY} e(y) \times F(x, y)}$
  and
  $F(x,y) = f(\elemOne_y)(x)$.

  This restricts to
  a
  $[0,1]$-linear Scott-continuous isomorphism
  \Eqn{e061444}
  given by
  taking duals $g \mapsto g^\intercal$
  \unskip.
  \unskip\footnote{
    The latter could be called
    an isomorphism of ``directed complete effect modules'' \cite{Jacobs:17}.
    See also the ``full Kegelspitzes'' (lit.\@ `cone tips') of \cite{Keimel:17}.
  }
  \begin{subequations}
  \begin{align}
    \label{e061445}
    \{F \In \XX \times \YY \Fun \RExtd\}
    &\cong
    \setcmp{f \In \Pred \YY \Fun \Pred \XX}{f \, \textnormal{linear, Scott-cts.\@}}
  \\
    \label{e061444}
    \{g \In \XX \Fun \DistPartial \YY\}
  &\cong
  \setcmp{f \In \Pred \YY \Fun \Pred \XX}{f \, \textnormal{linear, partial, Scott-cts.\@}}
  \end{align}
  \end{subequations}
\end{proposition}

\medskip
\noindent
Recall the
definition of the tensor
(p\pageref{paraTensor}):
\begin{equation}
  \label{e3424}
(f \otimes f')(\elemOne_{(z,w)})(x, y) = f(\elemOne_z)(x) \times f'(\elemOne_w)(y)
\end{equation}
By \Eqn{e061315}, $f \otimes f'$ is indeed uniquely defined,
with explicit formula
\begin{equation}
  \label{e101604}
(f \otimes f')(e)(z, w) = \sum_{x, y} f(\elemOne_x)(z) \times f'(\elemOne_{y})(w) \times e(x, y)
\end{equation}
It follows that $\otimes$ is functorial:
\begin{align}
\label{e1622a}
(\fnId_\XX \otimes \fnId_\YY)
  &=
  \fnId_{\XX \shorttimes \YY}
  \\
\label{e1622b}
(f \otimes g) \Comp (f' \otimes g')
  &=
(f \Comp f') \otimes (g \Comp g')
\end{align}
and indeed, $f \otimes g$ is bilinear and Scott-cts.\@ in its arguments.

\smallskip
Using the d-cone isomorphism
\Eqn{e061445}
between $\Pred \YY$ and the linear, Scott-cts.\@ functionals $\Pred \YY \Fun \RExtd$,
by abuse of notation
we may also take tensor products with $\Pred \YY$ itself
(e.g.\@ see Eqns.\@ (\ref{eWplUnvar},\ref{eWplPrint})).
Applying \Eqn{e101604} gives
\begin{align}
  \label{e1623a}
  (f \otimes g) (e \otimes e')
  &= f(e) \otimes g(e')
  \\
  \label{e1623b}
  (e \otimes \elemOne_\ZZ)
  &=
  (\fnId_\YY \otimes \elemOne_\ZZ)(e)
  = e^\ZZ
  \\
  \label{e1623c}
  \elemOne_\XX \otimes \elemOne_\YY
  &=
  \elemOne_{\XX \shorttimes \YY}
\end{align}

\begin{proposition}
  \label{p1628}
  Let $f \In \Pred \ZZ \Fun \Pred \XX, g \In \Pred \WW \Fun \Pred \YY$
  be linear, Scott-cts.\@ functions.
  If $f$ and $g$ are partial then $f \otimes g$ is partial.
\end{proposition}

\begin{proof}
  Suppose $f(\elemOne_\ZZ) \leq \elemOne_\XX$
  and $g(\elemOne_\WW) \leq \elemOne_\YY$.
  Then by monotonicity,
  $(f \otimes g) (\elemOne_{\ZZ \shorttimes \WW})
  =
  f(\elemOne_\ZZ) \otimes g(\elemOne_\WW)
  \leq
  \elemOne_\XX \otimes \elemOne_\YY
  =
  \elemOne_{\XX \shorttimes \YY}
  $.
\end{proof}

\subsection{Loss functions}
\label{sec860}

We state some additional
properties relating to partiality that are required for
\LEM{l101552}:
\begin{gather}
  \label{e1249}
  f(e_\YY) \leq e_\XX
  \quad\IMP\quad
  \fnMap(f)(\UpperCl e_\YY) \HRef \UpperCl e_\XX
\end{gather}

\begin{proposition}
  \label{p1250}
  If $f$ is a partial predicate transformer
  then
  $\fnMap(f)$ is partial.
\end{proposition}

\begin{proof}
  Follows directly from \Eqn{e1249},
  which in turn follows from $\UpperCl f(e_\YY) \supseteq \UpperCl e_\XX$.
\end{proof}

\begin{proposition}
  \label{p1750}
  Let
  $f_i \In \Loss \YY \Fun \Loss \XX$
  and
  $g_i \In \Pred \XX$
  for all $i \In I$.
  If all $f_i$ are partial
  and
  $\sum_{i \In I} g_i \leq \elemOne_\XX$
  then
  $\sum_{i \In I} g_i \boxtimes f_i(E)$
  is partial in $E$.
\end{proposition}

\begin{proof}
  $
  \sum_{i \In I} g_i \boxtimes f_i(\UpperCl \elemOne_\YY)
  \HRef
  \sum_{i \In I} g_i \boxtimes \UpperCl \elemOne_\XX
  =
  \sum_{i \In I} \UpperCl (g_i \boxtimes \elemOne_\XX)
  =
  \UpperCl \sum_{i \In I} g_i
  \HRef
  \UpperCl \elemOne_\XX
  $.
  The (in)equalities respectively follow from:
  \EQn{e1249};
  definition of $\fnMap$;
  linearity and Scott-cty.\@
  of the embedding $e \mapsto \UpperCl e$;
  and
  by assumption.
\end{proof}

\section{Example -- Rabin's mutex algorithm}
\label{sec153}

\citeauthor{lynch1991analysis}
\cite{lynch1991analysis}
find that
a randomised mutual exclusion algorithm
from
\cite{Rabin:82}
is unfair:
even under certain liveness assumptions,
the scheduler
can prevent a certain process
from acquiring the lock with
probability
$1 - o(c^{n})$
for some $c < 1$,
where $n$ is the number of processes.

The issue stems from a combination of factors,
including
a leakage of
hidden information
from processes
to the scheduler,
followed by
and the ability of the scheduler to
manipulate the
state of other processes
contending for the lock.
This suggests that our QIF-based approach might be applicable.
However it is not immediately clear how to model
the scheduler
and its liveness assumptions
using $\KuifjeNonDet$ (incl.\@ datatypes),
and we leave this as
a direction for future work.

The program \Eqn{e091510} models a highly simplified two-process concurrent system
based on a fragment of \cite{Rabin:82}.
At line 3, the nondeterministic scheduler may (or may not) allow thread 1 to enter and exit its critical section.
Then, both threads contend for the critical section simultaneously,
and this is resolved based on whether $\progVar{x} = \progVar{y}$.

Similarly to \EQn{e1833},
an adversarial scheduler with access to both variables
might choose to recompute $\progVar{x}$ at line 3 only when $\progVar{x} \neq \progVar{y}$.
Then the probability of thread 2 acquiring the lock is reduced to $\frac{1}{4}$.
\begin{gather}
\begin{lgathered}
  \progVar{x} \Assign 0 \PC{\frac{1}{2}} 1; \textit{// thread 1} \\
  \progVar{y} \Assign 0 \PC{\frac{1}{2}} 1; \textit{// thread 2} \\
  \textit{// assume thread 1 requests lock} \\
  \textit{// assume thread 2 requests lock} \\
  \Skip \;\sqcap\; (\progVar{x} \Assign 0 \PC{\frac{1}{2}} 1); \textit{// scheduler may run thread 1} \\
  \If \progVar{x} = \progVar{y} \; \{ \\
    \qquad \textit{// thread 1 acquires lock} \\
  \} \Else \{ \\
    \qquad \textit{// thread 2 acquires lock} \\
  \}
\end{lgathered}
\label{e091510}
\end{gather}

\fi

\end{document}